\documentclass[usenatbib]{mnras}

\usepackage{newtxtext,newtxmath}

\usepackage[T1]{fontenc}
\usepackage{ae,aecompl}

\usepackage{graphicx}	
\usepackage{amsmath}	
\usepackage{amssymb}	
\usepackage{aas_macros}
\usepackage{tabularx}
\usepackage[roman]{parnotes}
\newcommand{\cirx}{Cir X-1}

\title[Circinus X-1 jets]{The twisted jets of Circinus X-1}

\author[Coriat M. et al.]{
M. Coriat,$^{1,2,3}$\thanks{E-mail: mickael.coriat@irap.omp.eu}
R. Fender,$^{4}$
C. Tasse,$^{5,6}$
O. Smirnov,$^{6,3}$
A.K. Tzioumis,$^{7}$ 
 J.W. Broderick$^{8}$
\\
$^{1}$ IRAP, Université de Toulouse, CNRS, CNES, UPS, Toulouse, France\\
$^{2}$ Department of Astronomy, University of Cape Town, Private Bag X3, Rondebosch 7701, South Africa\\
$^{3}$ SKA South Africa, 3rd Floor, The Park, Park Road, Pinelands, 7405, South Africa\\
$^{4}$ Astrophysics, Department of Physics, University of Oxford, Keble Road, Oxford OX1 3RH, UK\\
$^{5}$ GEPI, Observatoire de Paris, CNRS, Université Paris Diderot, 5 place Jules Janssen, 92190 Meudon, France\\
$^{6}$ Department of Physics \& Electronics, Rhodes University, PO Box 94, Grahamstown, 6140, South Africa\\
$^{7}$ Australia Telescope National Facility, CSIRO, PO Box 76, Epping, New south Wales 1710, Australia\\
$^{8}$ ASTRON, the Netherlands Institute for Radio Astronomy, Postbus 2, 7990 AA Dwingeloo, The Netherlands\\
}

\date{Accepted XXX. Received YYY; in original form ZZZ}

\pubyear{2019}

\begin{document}
\label{firstpage}
\pagerange{\pageref{firstpage}--\pageref{lastpage}}
\maketitle

\begin{abstract}
We present the results of millimetre (33 and 35 GHz) and centimetre (2.1, 5.5 and 9.0 GHz) wavelength observations of the neutron star X-ray binary Circinus X-1, using the Australia Telescope Compact Array. We have used advanced calibration and deconvolution algorithms to overcome multiple issues due to intrinsic variability of the source and direction dependent effects. The resulting centimetre and millimetre radio maps show spatially resolved jet structures from sub-arcsecond to arcminute angular scales. They represent the most detailed investigation to date of the interaction of the relativistic jet from the X-ray binary with the young supernova remnant in which it is embedded. Comparison of projected jet axes at different wavelengths indicate significant rotation of the jet axis with increasing angular scale. This either suggests interactions of the jet material with surrounding media, creating bends in the jet flow path, or jet precession. We explore the latter hypothesis by successfully modelling the observed jet path using a kinematic jet model. If precession is the right interpretation and our modelling correct, the best fit parameters describe an accreting source with mildly relativistic ejecta ($v = 0.5 c$), inclined close to the plane of the sky ($i = 86\degr$) and precessing over a 5-year period. 

\end{abstract}




\section{Introduction}

Neutron star X-ray binaries (NSXBs) present us with a key opportunity to study radio jet behaviour in the absence of a black hole engine, allowing them to act as a control sample with which to study the possible effects the presence of event horizons, ergospheres and other black hole related properties can have on jet formation \citep{migliari06}. Unfortunately, neutron stars jets are more difficult to study as the source group tends to be more radio quiet than their black hole counterparts, especially at lower X-ray luminosities (\citealt{fender01a,migliari05}; \citealt*{gallo18}), making their radio emission more challenging to detect. However, with recent advancement in radio telescope arrays and data reduction techniques together with correct target selection this can eventually be overcome.

Circinus X-1 (Cir X-1) is a confirmed NSXB \citep{linares10} known for its regular 16.6 day flares (radio: \citealp{whelan77}, IR: \citealp{glass78}, X-ray: \citealp*{tennant86}), believed to be the result of an eccentric orbit ($e \sim 0.40-45$) \citep*{jonker07,johnston16} and increased accretion at periastron. Attempts to further classify the system in terms of atoll and Z source behaviour are especially difficult based on its unique X-ray behaviour, with indicators reminiscent of atoll, Z source \citep{oosterbroek95, shirey98} and even cases which fit neither during phases prior to periastron \citep{soleri09}. Cir X-1 also displays relativistic jets, resolved at a variety of scales and wavelengths (X-ray: \citealp{heinz07,soleri09a}, radio: \citealp{stewart93,fender98, tudose06}). In the past, Cir X-1's flares were found to precede brightening of nearby ejecta, which was interpreted as re-energisation via interaction with unseen outflows. The time delay between radio core flaring and re-brightening of the downstream material indicated very high Lorentz factors of $\Gamma$ $>$ 15 \citep{fender04a}, and while the estimates have been corroborated via further advanced analysis of the same data-sets \citep{tudose08}, subsequent observations of the source have yet to yield similar results \citep{calvelo12, miller-jones12}.

Cir X-1's quiescent and flare levels have varied significantly since discovery, with radio flares reaching $>$ 1 Jy in the late 1970s (\citealt{haynes78}; \citealt*{nicolson80}), but declining since $\sim$ 1997 to reach only 10s of mJy \citep{fender05}. Though inter-flare monitoring indicates the system remains historically `faint', in the past few years Cir X-1 has begun to brighten in the radio band. \citet{nicolson07} reported peak radio flares reaching Jansky-levels at 8.5 GHz for the first time in 20 years, in observations with the HartRAO 26-m telescope. Following the return to strong X-ray flaring in June 2010 \citep{nakajima10}, \citet{calvelo10} reported flares $>$ 0.1 Jy. The last reported radio observing campaign on Cir X-1 carried out in 2012 with the Karoo Array Telescope test array KAT-7, revealed strong radio flaring at Jy-levels \citep{armstrong13}. Long term X-ray monitoring of the system via the Monitor of All-sky X-ray Image \citep[MAXI:][]{matsuoka09} campaign has shown that Cir X-1 remains in a faint state for the majority of time, but is punctuated by short periods (weeks to months) of intense activity when both periastron flares and inter-flare levels become far brighter than average.

A striking feature of \cirx 's environnement is the large-scale radio nebula in which the binary is embedded. Historically thought to be produced by the jets
\citep{stewart93}, the nature of this nebula remained unclear until recent X-ray observations revealed an arcminute scale diffuse X-ray emission matching the structures observed in radio \citep{heinz13}. From morphological and spectral arguments the nebula was identified as the remnant of the supernova that gave birth to the neutron star. The properties of the remnant, together with a refined distance of 9.4 kpc obtained with X-ray dust scattering light echoes \citep{heinz15}, placed an upper limit of $t < 5400$ years on the age of the system. This upper limit makes \cirx\ the youngest known X-ray binary and an important test case for the study of both neutron star formation and orbital evolution in X-ray binaries. The youthfulness of the system could explain its complex behaviour observed on short and long timescales since the binary should still be settling down after the supernova explosion. 

The orientation and morphology of the jets are among the complex features of the system. Over the years, \cirx 's jets have been studied at various size scales from cm and mm wavelengths observations leading to conflicting results. \citet{calvelo12} reported on radio cm monitoring over a complete orbit of the binary and indicated that the system was behaving differently to what had been observed in the more active past. Lower levels of variability were found in structures around the system's core, the strongest of which were detected to the north-west: a region previously associated with the receding jet. Furthermore, the axis of near core resolved structure differs significantly from that previously observed, with the relatively similar intensity of the opposite components implying an inclination further from the line of sight. \citet{calvelo12a} reported on mm observations of \cirx , the first detection of a confirmed NSXB at mm wavelengths, and the subsequent image analysis provided evidence for sub-arcsecond jet structure around \cirx . This time, however, the structural axis indicated an angle closer to that of pre-2005 observations, albeit with the strongest signs of variability once again appearing to the north-west. Closer to the core, \citet{miller-jones12} reported on Australian Long Baseline Array (LBA) observations of \cirx\ showing a near east-west jet axis on milli-arcsecond scales, as well as indications of a higher (almost perpendicular to the line of sight) jet inclination angle, visibly different from the cm and mm images of  \citet{calvelo12, calvelo12a}. Together, these results suggest that either \cirx 's outflows have/are precessing or they have been deviated significantly from their former paths. Such a change could be recent, as the system showed little sign of precession over the previous decades of radio imaging \citep{tudose08}, though jet curvature has been visible on arcminute scales \citep[20 cm]{tudose06}.

Unfortunately, the spacing between the above observations makes determining the origin of the variable jet axis difficult. Depending on outflow velocities and the rate of any precession that is present, one may generate a spectrum of possible variable helical structures. Alternatively, we may be dealing with a new quasi-static flow path resulting from jet kinks and interaction with media of variable density within the nebula. Further constraints on the origin can be applied if we observe the system at multiple frequencies within a short period of time, thus allowing us to probe the system on a range of scales while minimising available time for structural variability to progress, and reveal any bends that may exist in the flows. This is the aim of the study presented in this paper. A description of the observations and the calibration process follows in \autoref{obs}. The radio maps obatined are analysed and discussed in \autoref{results} and concluding remarks are made in \autoref{sum}.

\section{Observations \& Data Processing}\label{obs}

\begin{table*}
\caption{For each observation the table lists the date, observing frequency, start Modified Julian Day (MJD), total on-source time before and after flagging, orbital phase estimated from \citet{nicolson07}, r.m.s. noise level of the image and measured flux density of the unresolved core component. We also indicate the bandpass and phase calibrators used. PKS 1934-638 was used for absolute flux calibration.}
\begin{center}
\begin{tabular}{lccccccccc}
\hline
\hline
Date & Frequency & MJD & On-source time (h)  & Orbital & r.m.s. & S$_{\nu}$ & Bandpass & Phase  \\
(UT) & (GHz) & start & [post-flag value (h)] & phase & ($\mu$Jy beam$^{-1}$) & (mJy) & calibrator & calibrator \\
\hline
2011 Dec 16 & 2.1 & 55910.75 & 9.67 [9.67]   & 0.67 & 50 & 12.80 $\pm$ 0.05 & PKS 1934-638 & PKS 1511-55 \\
2011 Dec 17 & 33 & 55911.82 & 5.77 [3.66]   & 0.73 & 35 & 0.78 $\pm$ 0.04 & PKS 1253-055 & PKS 1511-55 \\
2011 Dec 17 & 35 & 55911.82 & 5.77 [3.66]  & 0.73 & 35 & 0.68 $\pm$ 0.04 & PKS 1253-055 & PKS 1511-55 \\
2011 Dec 18 & 5.5 & 55912.74 & 9.58 [9.58]  & 0.80 & 10 & 4.65 $\pm$ 0.01 & PKS 1934-638 & PKS 1520-58 \\
2011 Dec 18 & 9.0 & 55912.74 & 9.58 [9.58]  & 0.80 & 12 & 3.11 $\pm$ 0.01 & PKS 1934-638 & PKS 1520-58\\

\hline
\end{tabular}
\end{center}
\label{table}
\end{table*}

We conducted a set of centimetre and millimetre observations of Cir X-1 over three consecutive days in 2011 Dec 16, 17 and 18 during a quiescent radio phase preceding a radio flare which started on December 22 \citep{armstrong13}. We observed the source using the Australia Telescope Compact Array (ATCA) in 6A configuration (minimum baseline of 337m, maximum of 5939m) at 2.1 GHz ($\sim$14 cm), 5.5 GHz ($\sim$6 cm), 9.0 GHz ($\sim$3 cm), 33 and 35 GHz ($\sim$8.8 mm). The observing log is given in \autoref{table}. Each frequency band was composed of 2048 $\times$ 1-MHz channels (making the 33 and 35 GHz bands contiguous). For all our observations we used PKS 1934-638 for absolute flux calibration. The sources used to calibrate the bandpass and the per-antenna complex gains as a function of time are listed in \autoref{table}. Flagging and initial calibration were carried out with the Multichannel Image Reconstruction, Image Analysis and Display (\textsc{miriad}) software \citep{sault95}. In the sections below we provide details on the additional data processing steps we applied on each dataset. Note that independently of the tools and methods we employed, we used Briggs' weighting with a robust parameter of 0.5 to produce the images presented in this paper.

\subsection{Millimetre data: 33 \& 35 GHz} \label{datared_34g}

\begin{figure}
\centerline{\includegraphics[width=0.49\textwidth]{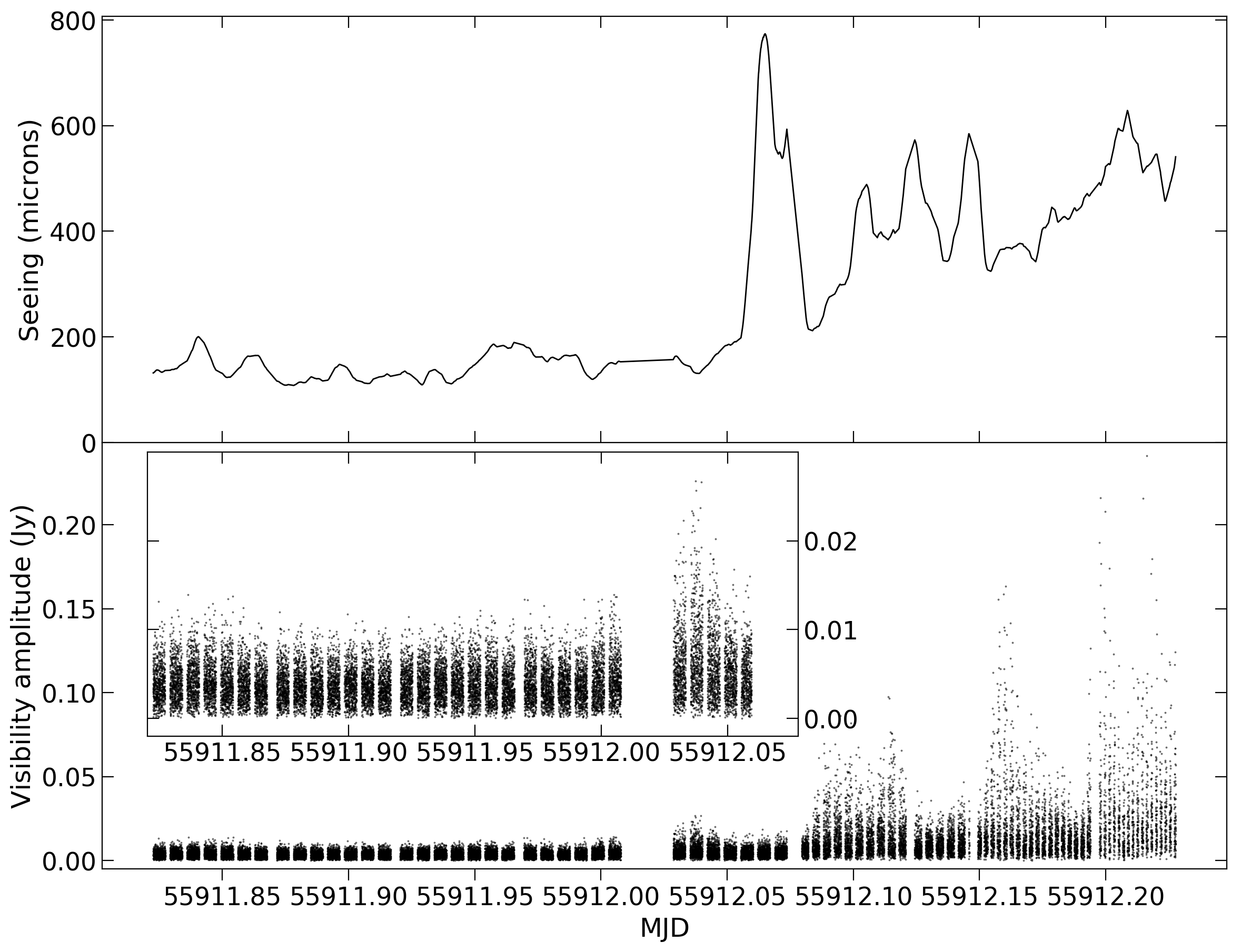}}
\caption{Cir X-1 seeing r.m.s. (top), and visibility amplitude (bottom) for the 2011 December 17 observation at mm wavelength  (34 GHz). The inset in the bottom panel shows the data fraction that was used to produce images and the source measurements listed in  \autoref{table}.}
\vspace{-10pt}
\label{cir-seeing}
\end{figure}

Since millimetre observations are more susceptible to atmospheric effects than those at longer wavelengths, phase stability had to be closely monitored via the ATCA ``seeing'' monitor (output gives r.m.s. path length fluctuations in microns; see \citealp{middelberg06}). These effects often increase as the runs progress, due to rising temperature and humidity. Upon analysis of the observations, it was found that large segments of data did not have sufficient phase stability to be reliable for imaging, making it necessary to define an adequate level of flagging that would improve final image fidelity. A maximum amount of decorrelation of 10 per cent arising from atmospheric seeing appears reasonable for the purpose of our work. For the 10 minutes calibrator/target cycle time we used, the corresponding r.m.s. of the path length fluctuations for the ATCA in 6A configuration at a wavelength of 8.8 mm, is $\sim 220$ microns\footnote{Using Eq. 2 and 6 from \citet{middelberg06} with a baseline length of 6 km and a Kolmogorov exponent $\beta/2 = 0.33$. Note that during the second half of the observation, we decreased the target/calibrator cycle time down to 5, 3 and finally 2 minutes to minimise decorrelation. However, we could not reach less than 10 per cent decorrelation according to the calibrator cycle time calculator \url{https://www.narrabri.atnf.csiro.au/calibrators/calcycle.html}.  }.  Segments whose majority of time was spent with r.m.s. higher than 220 microns were therefore excluded from subsequent analysis. \autoref{cir-seeing} shows the visibility amplitudes and path length r.m.s. in microns for the mm observation on Dec 17. The second half of the observation suffered from poor phase stability and has been excluded. The time segment that we used to filter the 33 GHz and 35 GHz datasets before imaging is shown in the inset of the bottom panel of  \autoref{cir-seeing}. Post-flagging on-source time is also given in \autoref{table}. 

Once the individual 33 and 35 GHz complex gains were calibrated using external calibrator sources in \textsc{miriad}, we converted the calibrated visibilities of the target field into measurement sets (MS) for further analysis with the Common Astronomy Software Application \citep[\textsc{casa},][]{mcmullin07}.  We applied a 32:1 frequency averaging to make the data volume more manageable. No time averaging was applied, so the standard 10s integration time was preserved. 

\begin{figure*}
\includegraphics[width=1.02\textwidth]{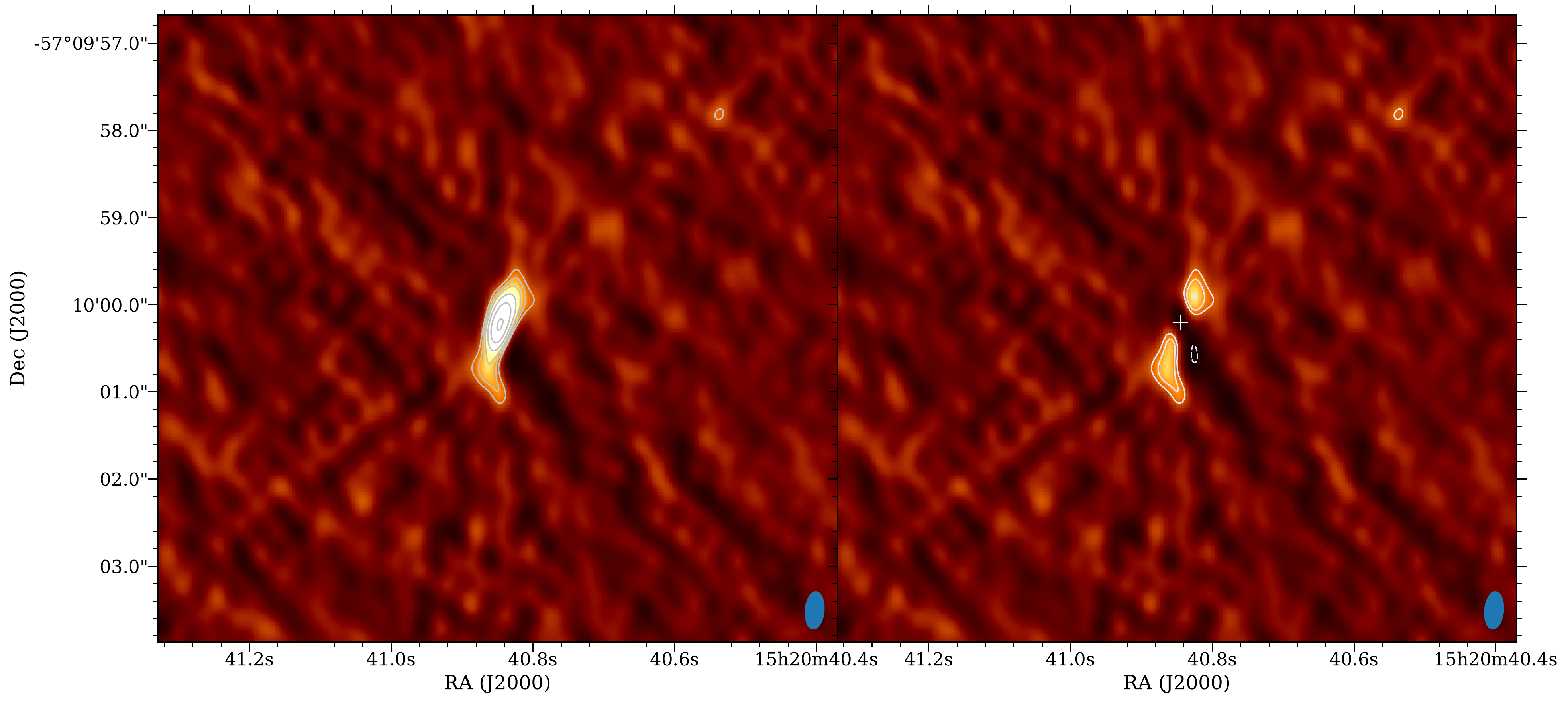}
\caption{Cir X-1 ATCA radio images with overlaid contours of the combined 33 and 35 GHz (central wavelength of $\sim$ 8.8 mm) measurement sets. The left panel shows the final map obtained after a self-calibration procedure (see text, section \ref{datared_34g}) . The right-hand map has had a fitted point source subtracted in the image plane at the location of the white cross, revealing symmetrical jet-like structures. The lowest contours correspond to $\pm 3 \sigma$ (where $\sigma = 35 \mu$Jy beam$^{-1}$ is the r.m.s. noise level). Contours are then increased by $(\sqrt 2)^{n} \sigma$, where $n \in [1,5]$. The synthesised beam size is $0.44 \times 0.23$ arcsec$^2$ in position angle $-5\degr.1$ and is shown at the bottom right-hand corner.}
\label{mm}
\end{figure*}

We produced the first images of the target field from the 33 and 35 GHz MS using straightforward Multi-Frequency Synthesis (MFS) deconvolution within the \textsc{clean} algorithm in \textsc{casa}. Cir X-1 is clearly detected at both frequency bands and is the only source visible in the field. The images revealed typical phase-error-related artefacts that we removed by performing three cycles of phase self-calibration using the \textsc{clean} components map as sky model. A final amplitude and phase self-calibration cycle was performed to remove weak residual artefacts. We finally combined the 33 and 35 GHz self-calibrated visibilities into a single MS which was imaged to form the final map shown in the left panel of \autoref{mm}. We subtracted in the image plane a fitted point source at the location of Cir X-1's core to better reveal the jet components. The residual image is shown in the right-hand panel of \autoref{mm}.

\subsection{Centimetre data: 5.5 \& 9.0 GHz} \label{datared_5g}

We applied to the 5.5 GHz dataset the same initial data reduction procedure as for the millimetre data. The first deconvolved images produced using MFS imaging showed significant radial artefacts affecting the central source Cir X-1. We performed a first standard phase self-calibration iteration with \textsc{casa};  our sky model was the CLEAN component map obtained from a shallow deconvolution (100 CLEAN iterations). The strongest artefacts were removed, but some artefacts were still present and could not be suppressed by further iterations of self-calibration. These remaining artefacts corrupted the source spatial structure and geometry and therefore had to be removed for a proper analysis of the jets' properties. A possible origin of these stubborn artefacts could be intrinsic variability of the source during the observing time. Indeed, one of the main assumptions in aperture synthesis interferometry is that the sky brightness distribution is constant during the time of observation. If the variable source is the only (or the dominant) source in the field, its variability will be absorbed by the self-calibration gain-amplitude solutions. In the presence of other sources, variability over the synthesis time produces artefacts that cannot be removed by neither a standard deconvolution algorithm nor by self-calibration.

A calibration scheme which can solve for these effects must be employed and for these observations, the differential gains algorithm \citep{smirnov11a} was used, implemented using the \textsc{calico} framework within the \textsc{meqtrees} software package \citep{noordam10}. The algorithm works by solving for additional
complex gain terms against an assumed sky model for a selected subset of sources. The differential gains approach is intended to deal with the more general case of Direction Dependent Effects (DDEs), but readily adapts itself to the case of variable sources. Since solving for direction-dependent gains increases the suppression 
of unmodeled sky sources (Nunhokee 2015, Master thesis\footnote{see \url{http://hdl.handle.net/10962/d1017900}}), we must be careful to restrict the number of degrees of freedom (DoF) in the solution. A feature of the \textsc{calico} implementation allows one to constrain the differential gain term to be identical across all antennas, which makes it effectively a proxy for source variability, with the minimum number of DoFs. The measurement equation then becomes:

\begin{equation}
\mathbfss{V}_{\mathit{pq}} = \mathbfss{G}_p ( \Delta\mathbfss{E} \, \mathbfss{X}_0 \, \Delta\mathbfss{E}^H + \mathbfss{X}_\mathrm{rest}) \mathbfss{G}_q^H,
\end{equation}
where $\mathbfss{V}_{\mathit{pq}}$ is the visibility matrix measured by the interferometer formed by antennas $p$ and $q$,  $\mathbfss{X}_0$ are the model visibilities of the assumed variable source and $\mathbfss{X}_\mathrm{rest}$ are the model visibilities representing the rest of the local sky. $H$ represents the Hermitian transpose. Both models need a non-negligible amount of flux in order for the global gains $\mathbfss{G}_p$ and the differential gain $\Delta \mathbfss{E}$ to be independently constrained.

\begin{figure*}
\includegraphics[width=1.0\textwidth]{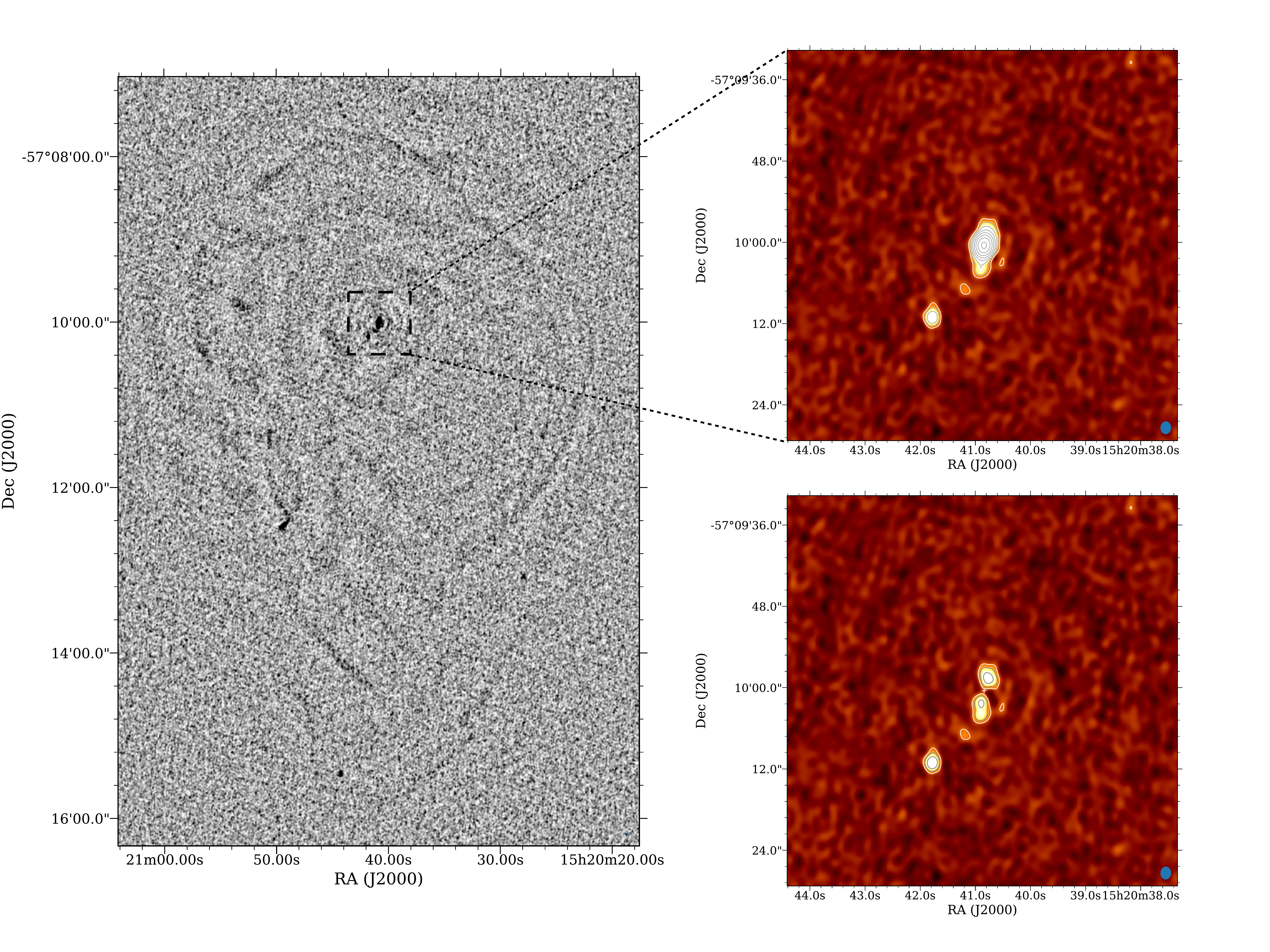}
\caption{Cir X-1 ATCA radio images at 5.5 GHz. The left panel shows the full size map obtained after self-calibration with \textsc{meqtrees} using the differential gains algorithm to absorb intrinsic variability. We can see \cirx\ at the centre surrounded by its nebula. High contrast greyscale has been chosen here to accentuate the (otherwise weak) structures of the nebula. The top-right panel is a close-up view of the central region showing \cirx\ and a weak source at the south-east which is presumably a background source. The bottom-right panel shows the same close-up where the central variable source has been subtracted to unveil the jets. The white cross indicates the location of the subtracted source. The contour levels in the right panel maps begin at $3 \sigma$ (where $\sigma = 10 \mu$Jy beam$^{-1}$  is the r.m.s. noise level) and increase in multiple of 2. A single, dashed negative contour is included with a value of $- 3 \sigma$. The synthesised beam size is $1.9 \times 1.6$ arcsec$^2$ in position angle $-3\degr.7$ and is shown at the bottom right-hand corner.}
\label{5g}
\end{figure*}

From the best image obtained with \textsc{casa}, we created an initial sky model using \textsc{pybdsf} \citep{mohan15} and \textsc{tigger}\footnote{\url{https://github.com/ska-sa/tigger}} where \cirx\ was initially modelled by a single point source and the rest of the sources in the field by gaussian or point sources. We then self-calibrated the data against this sky model calculating $\mathbfss{G}_p$ as a function of time and frequency, using \textsc{calico}'s smooth gain solver mode \citep[see][for a description]{smirnov15}, which effectively corresponds to doing a weighted-least square solution centred at each time/frequency point, with a Gaussian weighting kernel of 100 seconds and 640 MHz respectively. We then computed the corrected residuals, i.e., the residual visibilities obtained after subtraction of the model and application of the gain corrections. Corrected residuals allow for a quick evaluation of the quality of the calibration. If the corrected residuals 
are noise-like then the combination of the sky model and gain corrections correctly represents the data. 
The first corrected residuals we obtained showed symmetrical structures around the subtracted central point source as well as ring like artefacts. We then attempted to account for variability, by solving for $\Delta \mathbfss{E}$ with a Gaussian weighting kernel of 15 min (and no variation in frequency), and we kept the same time/frequency smoothing kernels as before for the $\mathbfss{G}_p$ solution. The corrected residuals we obtain after one calibration cycle were improved but the symmetrical structures were still present together with a negative hole at the location of the subtracted point source. This indicated that a single variable point source was not a good representation of the data. Therefore, we added two point sources to the sky model at the locations of the residual structures to account for the potential presence of jets. We first calibrated against this new sky model without the differential gain term and allowing the positions of the two point sources to vary from one calibration cycle to another. The symmetrical structures were finally removed but the residuals at the location of the central source were still unsatisfactory. We then added back the differential gain term for the central point source and performed another cycle of self-calibration/imaging which finally yielded noise-like residuals. \autoref{5g} shows the final image (where we can distinguish the nebula surrounding \cirx ) together with a close-up view of the central region where the core component has been subtracted to emphasise the jets.

\begin{figure*}
\includegraphics[width=1.0\textwidth]{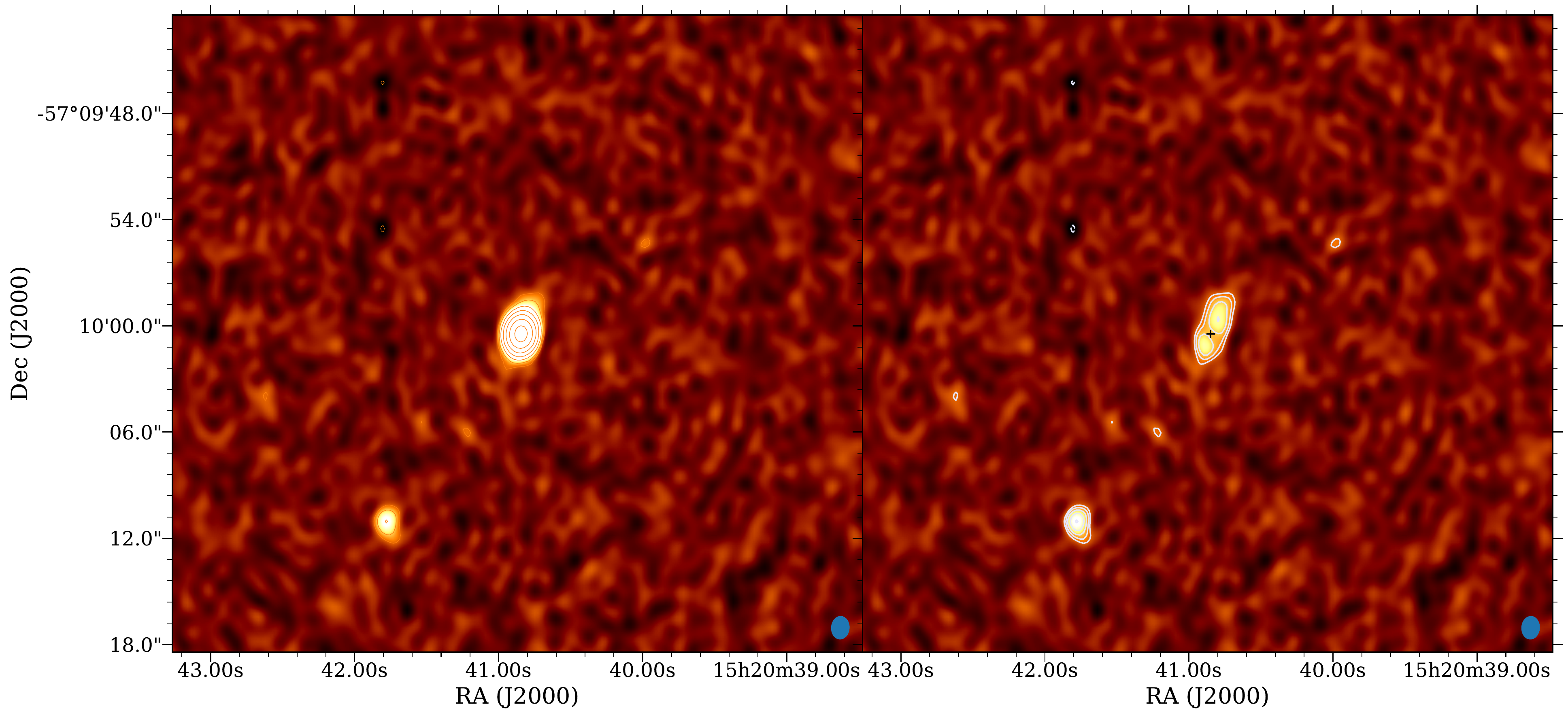}
\caption{Cir X-1 ATCA radio images at 9.0 GHz. The left panel shows the map obtained after self-calibration with \textsc{meqtrees} using the differential gains algorithm to absorb intrinsic variability. The weak (presumably background) source visible in the 5.5 GHz map to the south-east of \cirx\ is also detected in this map. The right panel shows the image where the central variable source has been subtracted to reveal the jets. The black cross indicates the location of the subtracted source. The lowest contours correspond to $\pm 3 \sigma$ (where $\sigma = 12 \mu$Jy beam$^{-1}$ is the r.m.s. noise level). Contours are then increased by $(\sqrt 2)^{n} \sigma$, where $n \in [1,5]$ for the right panel and only in multiples of 2 for the left panel for clarity. The synthesised beam size is $1.3 \times 1.1$ arcsec$^2$ in position angle $-3\degr.8$ and is shown at the bottom right-hand corner.}
\label{9g}
\end{figure*}

The 5.5 and 9.0 GHz data were obtained simultaneously thanks to the dual-frequency observation capabilities of the ATCA. Unsurprisingly, the 9.0 GHz images suffered from the same kind of corruption as the 5.5 GHz data. We thus employed the same calibration strategy and obtained the radio maps presented in \autoref{9g}. 

\subsection{Centimetre data: 2.1 GHz}\label{sec:2.1G}

\begin{figure*}
\includegraphics[width=0.74\textwidth]{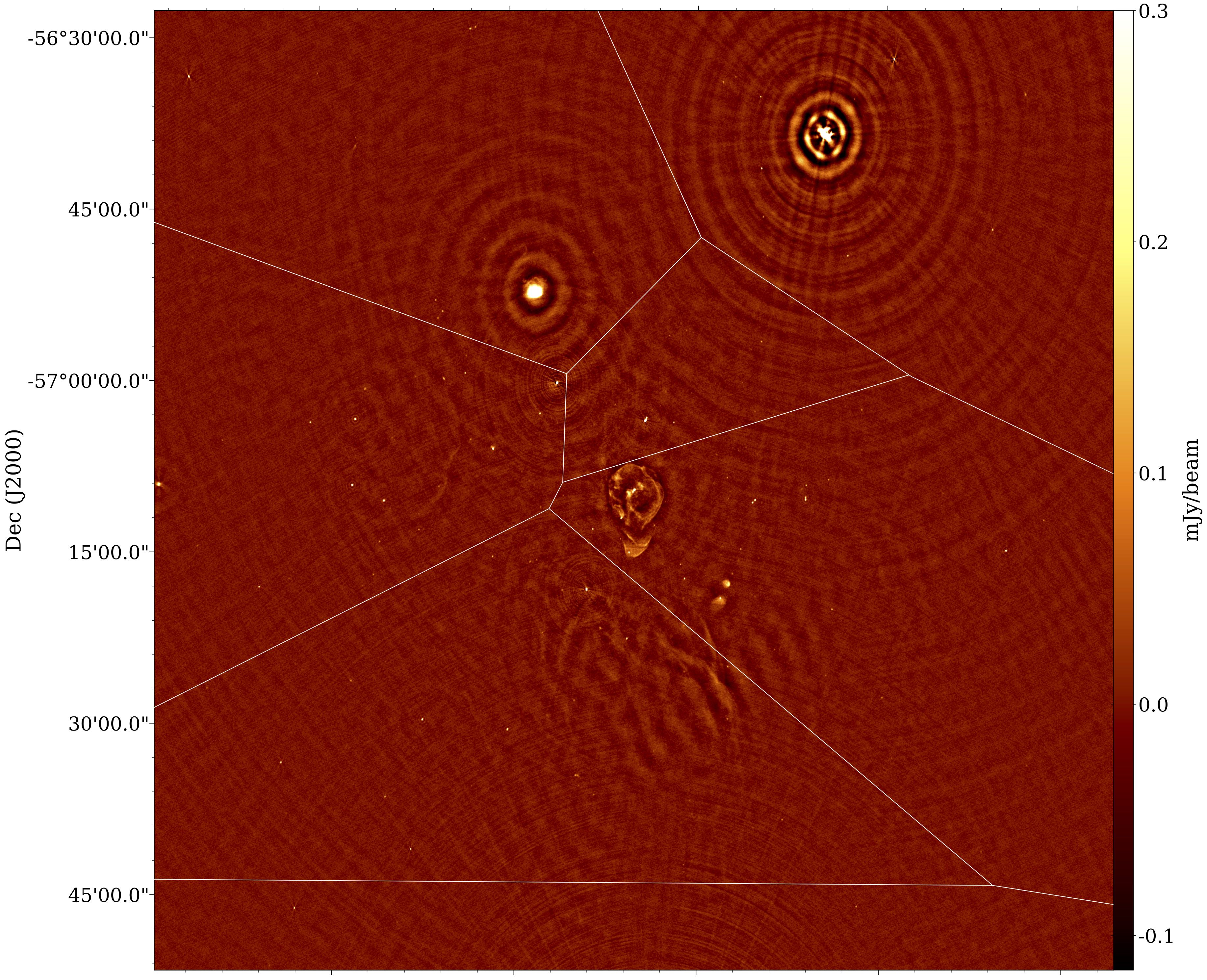}
\includegraphics[width=0.74\textwidth]{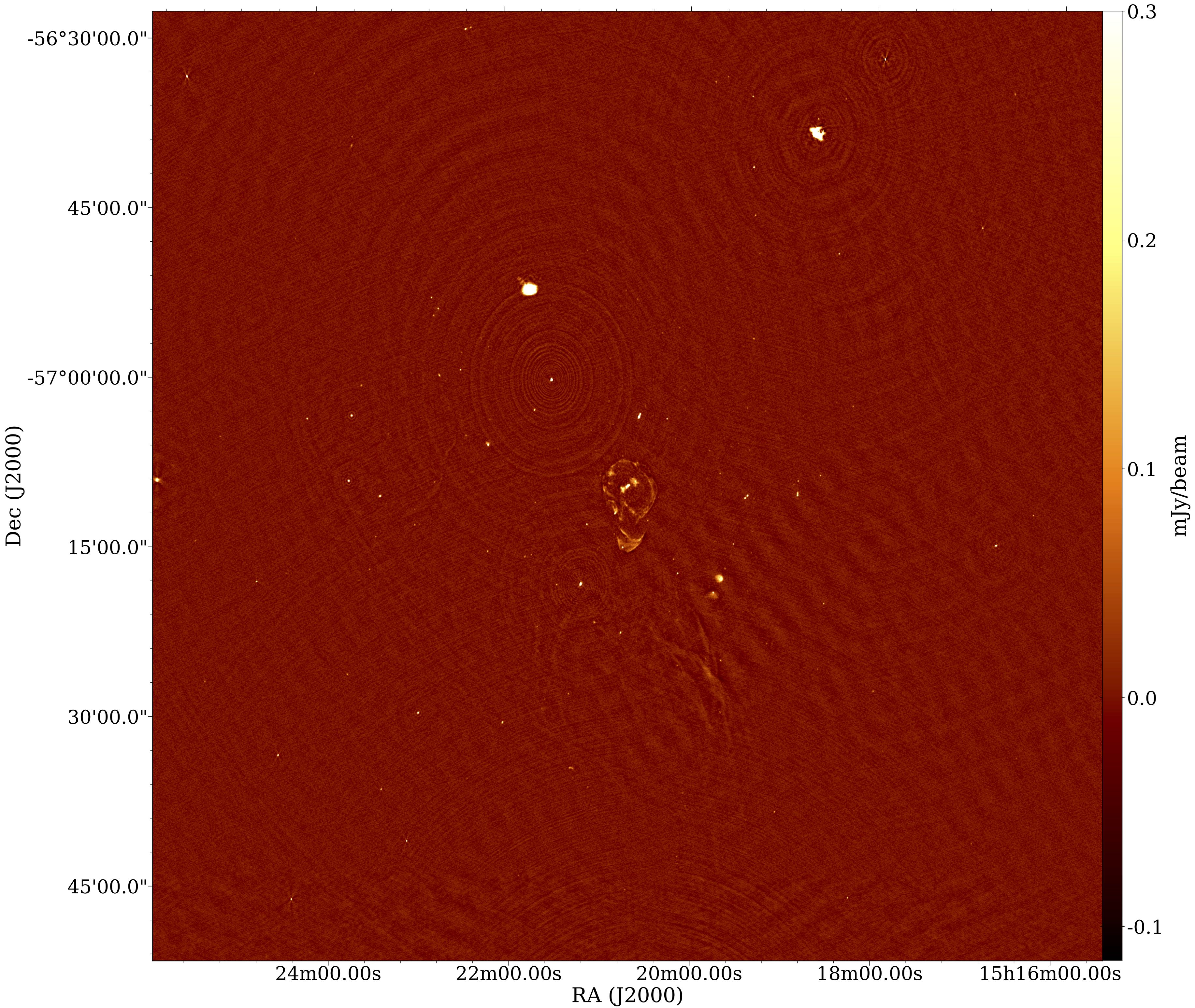}
\caption{ATCA large scale radio images at 2.1 GHz of \cirx's field. Top panel: Best image obtained after standard self-calibration and imaging. The colour bar shows the radio brightness in units of mJy~beam$^{-1}$.  Bottom panel: Image obtained after direction dependent calibration and imaging using \textsc{killms} and \textsc{ddfacet}. The polygons overlaid on the top panel corresponds to the 7 regions for which independent calibration solutions have been computed (see Section \ref{sec:2.1G}). The southern-most region is only partially displayed as we only show the central two third of the full-size image. \cirx\ and its nebula are located at the centre of the image. The field contains a set moderately strong and morphologically complex sources within and beyond the primary beam FWHM ($\sim 30\arcmin$) as well as faint and extended structures.}
\label{2g}
\end{figure*}

As we move to lower observing frequencies, the field of emitting sources surrounding \cirx\ becomes increasingly complex. At 2.1 GHz, this field contains various type of sources ranging from strong and compact to faint and extended (e.g. \cirx 's nebula) including partially extended, moderately bright, morphologically complex sources (to which \cirx\ belongs). Preliminary images also revealed the need of imaging well outside the primary beam FWHM ($\sim 30\arcmin$) due to the presence of strong peripheral sources producing sidelobes encroaching into the central part of the image. Deconvolution and imaging of such a field then becomes a challenge as direction dependent effects come into play adding to the intrinsic complexity of the source population. The best image we could obtain using standard imaging methods and direction independent self-calibration is presented in the top panel of \autoref{2g}. Numerous circular and radial artefacts are visible. This makes the field truly DDE limited (unlike the 5.5 and 9.0 GHz case, where only intrinsic source variability was of concern). We therefore took a different approach to the calibration, employing the new \textsc{killms}\footnote{\url{https://github.com/saopicc/killMS}} calibration package combined with the \textsc{ddfacet}\footnote{\url{https://github.com/saopicc/DDFacet}} imager \citep{tasse18}.  We provide below a brief description of these new software packages. 

The new generation of radio interferometers (SKA precursors and pathfinders as well as the upgrades of the current radio observatories) are characterised by wide fields of view, large fractional bandwidth, high sensitivity and high resolution. These superior performances introduce new issues for calibration and imaging due to complex signal corrupting effects that were negligible with the older generation of instruments. These effects can be of instrumental (e.g. pointing errors, dish deformation, antenna coupling within phased arrays) or astrophysical origin (e.g. the ionosphere and its associated Faraday rotation) and are often characterised by levels of corruption varying accros the field of view (being as well baseline, time and frequency dependent). Correcting these direction-dependent effects is a complex and computationally expensive process as it consists in inverting a very large and often ill-conditioned system of non-linear equations. \textsc{killms} implements two very efficient algorithms for solving this direction-dependent calibration problem. The \textsc{CohJones} and \textsc{KAFCA} solvers use optimisation techniques based on the properties of the complex (``Wirtinger'') Jacobian. The mathematical framework and related algorithmic implementation is described in \citet{tasse14} and \citet{smirnov15}. 

\textsc{ddfacet} is a wide-band wide-field imager based on a co-planar faceting scheme \citep{tasse18}. It implements new deconvolution algorithms and can account for externally defined gain corrections and beam patterns. It was designed to work in concert with \textsc{killms}. In each of the defined direction (i.e. region; see an example in \autoref{2g} top panel), a different set of Jones matrices is taken into account by \textsc{ddfacet} during the deconvolution cycle. Here, since the ATCA primary beam at 2.1 GHz is poorly constrained beyond the first minimum\footnote{see \url{http://www.narrabri.atnf.csiro.au/people/ste616/beamshapes/beamshape_16cm.html}}, we simply work with apparent flux densities and the Jones matrices being considered are the calibration solutions produced by \textsc{killms} (one scalar solution per direction, antenna, time and frequency).

To obtain the final image presented in the bottom panel of \autoref{2g}, we applied the following calibration and imaging procedure: we first produced an image with \textsc{ddfacet} using direction independent faceting and the wide-band Hybrid Matching Pursuit (HMP) deconvolution algorithm\footnote{An improved analog of the multi-scale multi-frequency CLEAN algorithm. See \citet{tasse18} for details.}. From this preliminary image we made a sky model and clustered the sky in 7 directions corresponding to the directions of the 7 brightest sources in the field. This decomposition in 7 regions is illustrated by the polygons shown in the top panel of \autoref{2g}.  We also created a deconvolution mask using a threshold-based algorithm combined with manual inclusions/exclusions of specific regions of the sky, e.g., excluding strong artefacts and including faint and extended sources. From the sky model and the cluster nodes catalog, we computed antenna-time-frequency-direction dependent gain corrections using the \textsc{CohJones} algorithm within \textsc{killms}. For this first pass we used time and frequency solution intervals of 10 min and 256 MHz respectively. The obtained calibration solutions were applied during the subsequent deconvolution and imaging step with \textsc{ddfacet}. With the created mask, we then used the Sub-Space Deconvolution (SSD) algorithm in \textsc{ddfacet} instead of HMP as it improves deconvolution of extended emission. We repeated this calibration/imaging cycle twice, improving sky model and mask at each round and reducing the frequency solution interval down to 64 MHz (we kept the same time bin of 10 min as lower values provided poorer results). Most of the artefacts have disappeared in the final image shown in the bottom panel of \autoref{2g}. Some weak residual structures remain at the noise level around a few sources but they do not affect the central region were \cirx\ is located. 

If \cirx\ was variable during the course of the 2.1 GHz observation, the facet containing it should have shown similar issues as the 5.5 and 9.0 GHz images. Yet, a single gain solution for the entire facet was enough to remove the artefacts which indicates that the source was likely not (or weakly) variable. This is consistent with the time sequence of our observations. We first observed at 2.1 GHz on Dec. 16 (orbital phase 0.67), then at 33 and 35 GHz on Dec. 17 (orbital phase 0.73) and finally at 5.5 and 9.0 GHz on Dec. 18 (orbital phase 0.80). The later observation was the closest in time to the radio flare detected on Dec. 22 which might explain why it is the only observation where variability was significant.

\begin{figure*}
\includegraphics[width=1.0\textwidth]{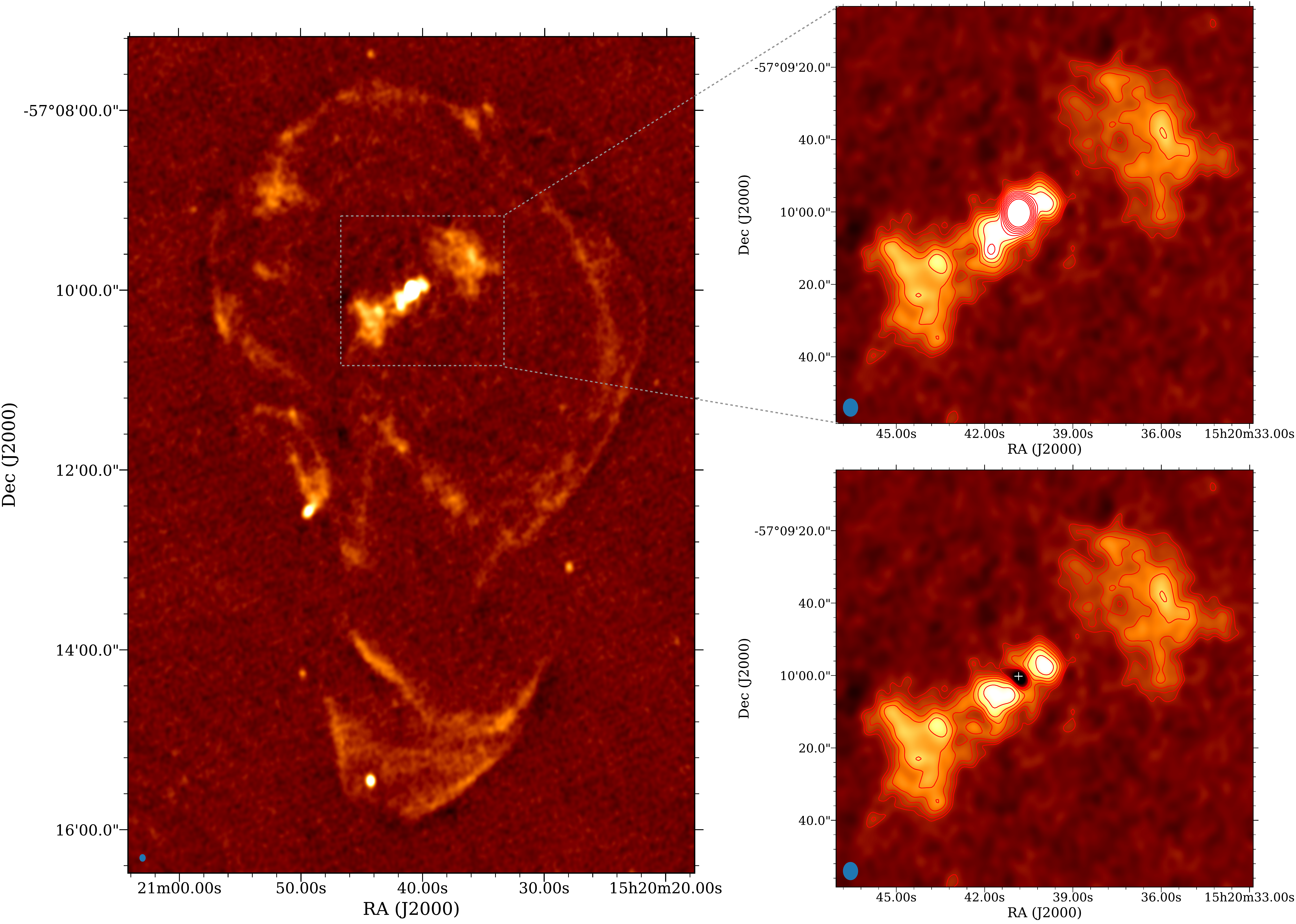}
\caption{Cir X-1 ATCA radio images at 2.1 GHz. The left panel shows the supernova remnant surrounding \cirx . The top-right panel is a close-up view of the central region where we clearly detect \cirx\ together with extended jet structures. The two lobes to the SE and NW (also detected in X-rays) are likely due to the interaction of the jets with surrounding media within the nebula. Contours are overlaid and start at $\pm 3 \sigma$ (where $\sigma = 14 \mu$Jy beam$^{-1}$ is the r.m.s. noise level) and increase by $(\sqrt 2)^{n} \sigma$, where $n \in [1,7]$. The background source to the SE detected in the 5.5 and 9.0 GHz images is also visible although it is partially blended with the surrounding jet emission.The bottom-right panel shows the same image where we have subtracted, in the image plane, the core source (location marked by the white cross) as well as the SE background source for clarity. The synthesised beam size is $5.0 \times 4.2$ arcsec$^2$ in position angle $1\degr.4$ and is shown at the bottom left-hand corner.}
\label{2g-closeup}
\end{figure*}

\autoref{2g-closeup} (left panel) shows a close-up view of this central region where the remnant surrounding \cirx\ is now clearly visible. Zoomed-in views of \cirx\ itself and its jets are shown in the right panels of \autoref{2g-closeup}. As for the datasets at higher frequencies we subtracted the contribution of the unresolved core radio emission to focus on the double-sided jets structure. Note that the scale of these structures are the largest on which we can be confident that the jets dominate over the remnant in terms of morphology.

\section{Results \& Discussion}\label{results}

\subsection{Jet flux and morphology}

\begin{figure}
\includegraphics[width=0.48\textwidth]{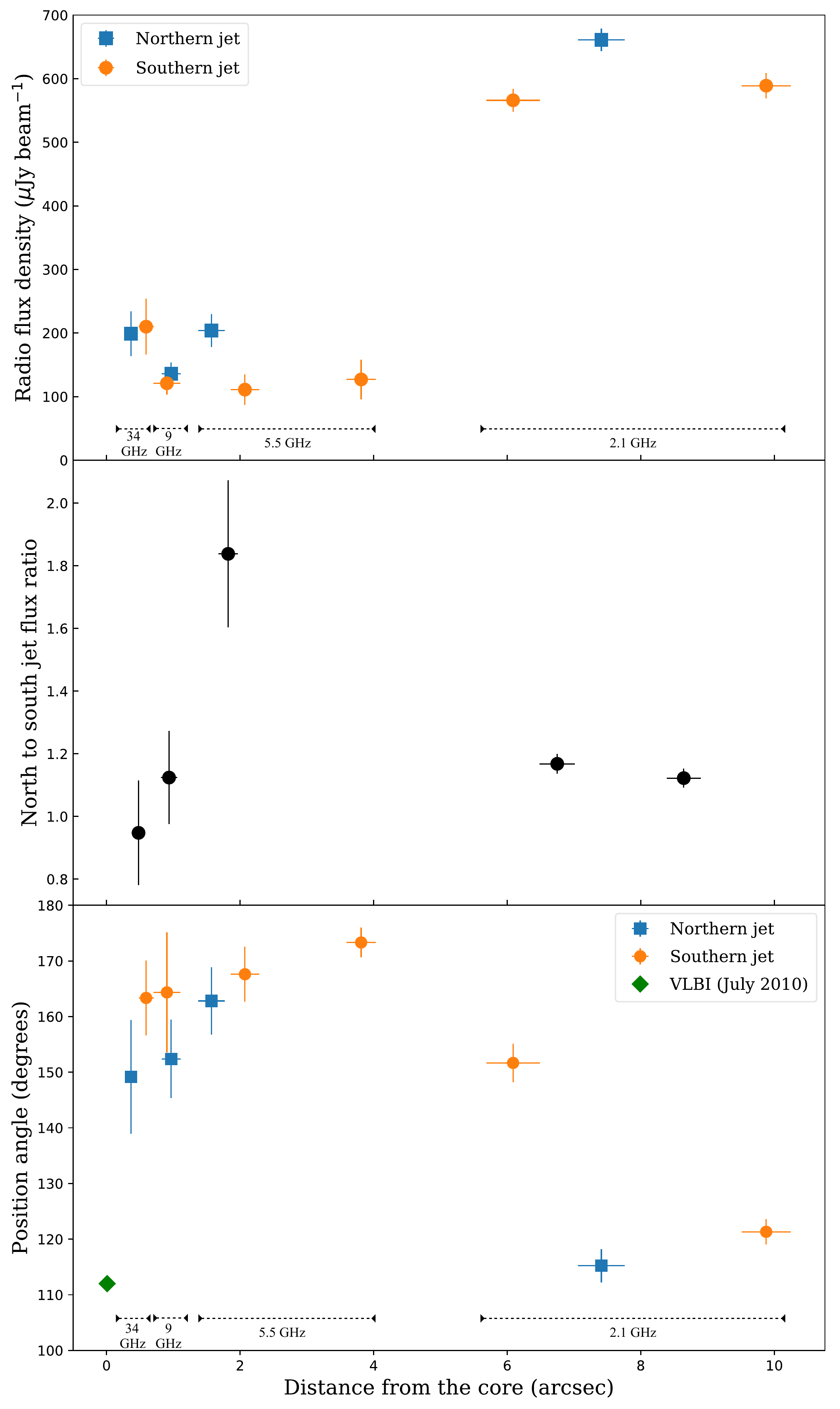} 
\caption{Radio flux densities (top panel), north to south jet flux ratio (middle panel) and jet position angle (bottom panel) as a function of distance to the core. Position and fluxes were obtained by fitting point sources to the northern and southern jet components in the image plane.  Error on the positions are estimated as the PSF size divided by twice the local signal-to-noise ratio \citep[see e.g.][]{fomalont99}. The bottom panel also includes the position angle of the jet measured at milliarcsecond scale in July 2010 by \citet{miller-jones12}. Note that the jet flux ratios are plotted against the mean distance of each component pairs.}
\label{fra}
\end{figure}

After subtraction of the core emission as described in the previous section, we measured the flux density and position of the jet components for each frequency band by fitting point sources to the image plane. The majority of the jet components being unresolved, only a single point source per  component was necessary. However, the southern components of the 5 GHz and 2.1 GHz jets are both slightly resolved. Two point sources were required to model each one of them. The top and middle panel of \autoref{fra} shows the jets flux density as well as the south to north jet flux ratio as a function of projected distance from the core\footnote{We do not present the flux densities in the form of a spectrum as the jet components are not spatially coincident.}, where we used the coordinates of the 34 GHz core source as reference. The southern jet components of the 5.5 and 2.1 GHz being modelled by two point sources each while their corresponding northern components comprise a single point source only, we calculated two flux ratios for each dataset and plotted them against the distance of the southern components

\begin{figure*}
\includegraphics[width=1.01\textwidth]{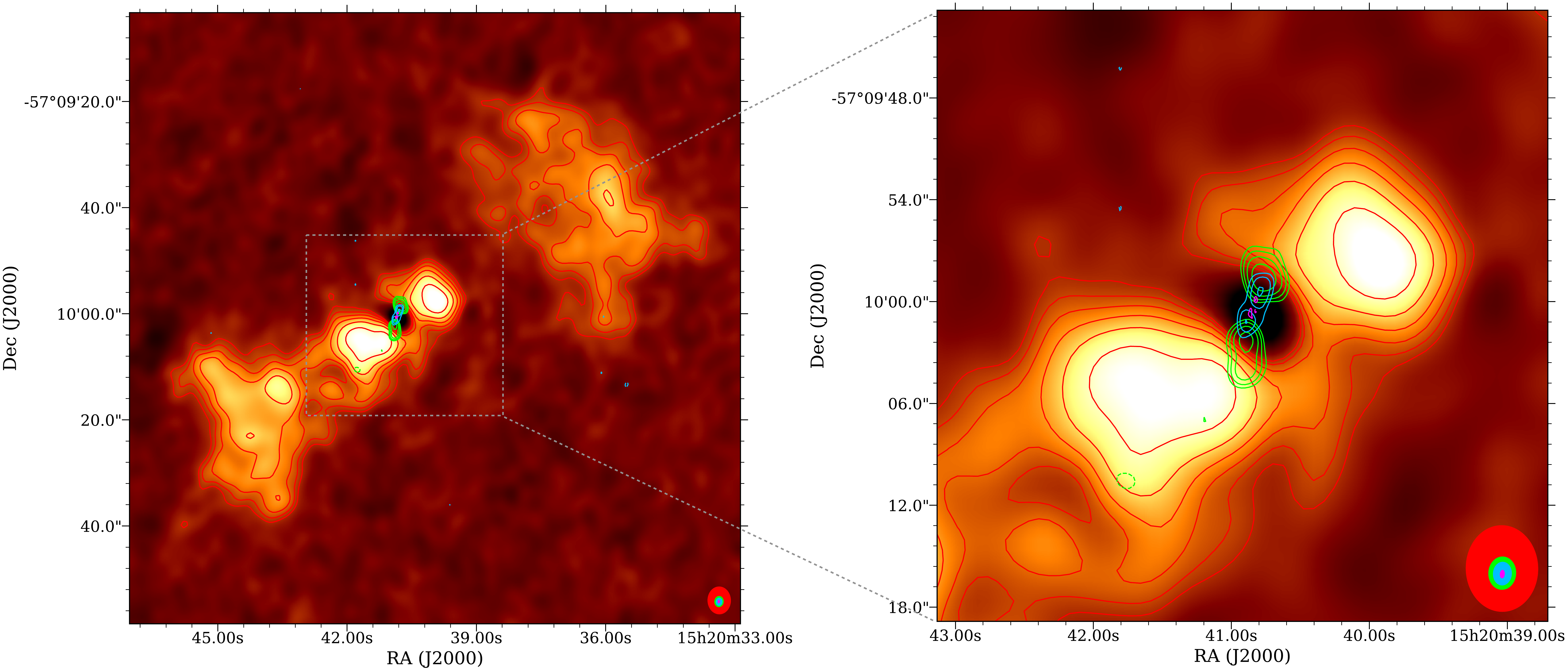} 
\caption{Central region of the 2.1 GHz image of \cirx\ with overlaid contours (listed in order of decreasing distance to the core) at 2.1 GHz (red), 5.5 GHz (green), 9.0 GHz (blue) and 34 GHz (magenta). For each frequency band, the central core source has been removed to focus on the jet morphology. The contour levels are the same as those presented in \autoref{mm}, \autoref{5g}, \autoref{9g} and \autoref{2g-closeup}. The synthesised beams corresponding to each observing frequency are shown at the bottom right-hand corner of the images with the same colour code as the one used for the contours and with size decreasing with increasing frequency. }
\label{combined}
\end{figure*}

From this figure we note that, on average, the northern jet is slightly brighter than its southern counterpart albeit the flux ratio does not exceeds two. This first excludes the possibility of a highly relativistic jet seen at low inclination with respect to the line of sight. It also suggests that the northern jet is the approaching one. In the bottom panel of \autoref{fra}, we plot the jet axis position angle as a function of distance to the core. For comparison, we also include the jet position angle measured at milliarcsecond scale with the LBA in July 2010 (17 months before our observations) by \citet{miller-jones12}.  A more qualitative view of the jet morphology is presented in \autoref{combined} where we overlay the jet contours at 5.5 GHz, 9.0 GHz and 34 GHz on the 2.1 GHz image. Both figures indicate significant variations of the jets orientation with increasing distance from the core. This flow path could obviously result from jet kinks and/or interactions with media of variable density within the nebula. Alternatively, a precessing jet may also produce a similar pattern. The later interpretation would also naturally explain the fact that the positions of the jet components  change symmetrically with respect to the centre when moving away from the core. In addition, we note from \autoref{fra} that the jet axis does not vary monotonically. Its position angle first increases then decreases with increasing distance from the core, suggesting a form of oscillation. Finally, the angular extent of the radio lobes located at $\sim 40 \arcsec$ from the core \citep[see \autoref{combined} and][for the X-ray counterpart]{sell10,heinz13} also suggests precession if we assume the lobes originate from interaction of the jets with the nebula. We thus explore further this scenario in the section below.

\subsection{Jet precession modelling}

In order to evaluate if a precessing jet could reproduce our data, we adapted the kinematic jet model of \citet{hjellming81a} initially developed to explain the jet morphology of the well-known microquasar SS433. The details of the model can be found in the aforementioned article. We recall below the main parameters:

\begin{description}
\item $i$: inclination of the jet precession axis with respect to the line of sight.
\item $\psi$: half opening angle of the jet precession cone.
\item $\chi$: position angle of the jet precession axis in the plane of the sky (positive from north to east).
\item $P$: precession period.
\item $v$: jet velocity. The model assumes constant velocity.
\item $d$: distance to the source.
\item $\phi_{\rm prec}$: precession phase.
\item $s_{\rm jet}$: jet sign parameter. $+1$ for the approaching jet, $-1$ for the receding jet.
\item $s_{\rm rot}$: sense of rotation. $+1$ corresponds to counterclockwise, $-1$ to clockwise.
\end{description}

For a given time $t_0$, the model computes the position of a particular twin jet pair ejected at time $t_{\rm ej}$, in the form of right ascension and declination offsets from the core position, $\Delta \alpha_{\rm mod}$ and $\Delta \delta_{\rm mod}$, respectively. Considering all the possible values of $t_{\rm ej} < t_0$, we obtain a snapshot of the model jet at time $t_0$, projected on the plane of the sky. The fitting procedure then requires an objective function to be minimised and that describes the quality of the fit. We require that a good solution be one that minimises the sum of the minimum distances of the data points  to the model in the ($\Delta \alpha$, $\Delta \delta$) plane. Ideally, any data point would sit on the model curve and the minimal distances would be zero for a perfect solution. Thus, we define our objective function as follow:
\begin{equation}
S = \Upsilon(\beta, i) +  \sum_{i}  \left[\frac{(\Delta \alpha_i - \Delta \alpha_{\text{mod}_i})^2}{\sigma_{\alpha_i}^2} + \frac{(\Delta \delta_i - \Delta \delta_{\text{mod}_i})^2}{\sigma_{\delta_i}^2} \right],
\end{equation}
where $\Delta \alpha_i$ and $\Delta \delta_i$ are, respectively, the right ascension and declination offset of the $i$th data point with associated errors $\sigma_{\alpha_i}$ and $\sigma_{\delta_i}$. The
chosen model ($\Delta \alpha_{\text{mod}_i}, \Delta \delta_{\text{mod}_i}$) pair is the one that minimises the square distance between the precession helix and the observed ($\Delta \alpha_i , \Delta \delta_i$) pair. To guarantee that our jet-precession modelling will also predict velocities and inclinations in agreement with the observed jet flux ratio, we included a penalty function  $\Upsilon(\beta, i)$ based on the analytical expression of the approaching to receding jet flux ratio $\frac{F_{\rm app}}{F_{\rm rec}}$:

\begin{equation}     
  \Upsilon(\beta,i) = \left\{\begin{array}{rl}
0,&\mbox{if}\quad  0.8 < \frac{F_{\rm app}}{F_{\rm rec}} = \left( \frac{1+ \beta \cos i}{1-\beta \cos i} \right)^{n} < 2 \, ,\\
100,&\mbox{elsewhere},
\end{array}\right.
\end{equation}
where $\beta = \frac{v}{c}$ is the velocity (in units of the speed of light) and $n$ depends on the geometry and spectral index of the jet and is typically in the range between 2 and 4 (we have used an intermediate value of 3). The choice of a penalty $\Upsilon(\beta,i) = 100$ effectively guarantees that any tentative solution providing $\frac{F_{\rm app}}{F_{\rm rec}} > 2$ or $<0.8$ (values based on the measured flux ratios, see \autoref{fra}) is statistically disfavoured during optimisation. Here we simply set bounds to the jet flux ratios as we do for the other parameters (see below). 

Regarding the optimisation process itself,  preliminary tests using standard gradient-based algorithms indicated that the objective function had  many local minima in which the algorithm easily got trapped. We therefore adopted a global optimisation approach more suited to our problem. We used the `differential evolution' algorithm \citep{storn97} implemented in the \textsc{scipy.optimize} package. The algorithm requires bounds for each parameter; we therefore defined the parameter space as follows:
\begin{description}
\item $i$: from $0\degr$ to $90\degr$  
\item $\psi$: we constrained the half opening angle of the jet precession cone to remain in the range $20\degr$ to $50\degr$  in order to match the half opening angle of $35\degr$ of the extended lobe structures that we assume result from interactions of the jets with the nebula \citep[see e.g.][]{sell10}. 
\item $\chi$: similarly, the position angle of the jet precession axis is restricted to the $110\degr$ to $140\degr$ range to match the position angle of the lobes.
\item $v$: from $10^{-2} c$ to $0.99 c$
\item $P$: from 10 to 3000 days. Preliminary tests have shown that outside of these boundaries (and for the velocity range adopted above) the model tends to either a straight line (P>3000 days) or a helix that entirely fills the space.
\item $d$: we fixed the distance to 9.4 kpc based on \citet{heinz15}.
\item $\phi_{\rm prec}$: from 0 to $2\pi$
\item $s_{\rm jet}$: fixed to $-1$ for the southern jet (receding) and $+1$ for the northern jet (approaching).
\item $s_{\rm rot}$: fixed for a given optimisation run. Both $+1$ (counterclockwise) and $-1$ (clockwise) were tested.
\end{description}

\begin{table}
\caption{Best-fit parameters of the precessing jet model applied to \cirx's radio jets.}          
\label{param}      
\centering                        
\begin{tabular}{lc}     
\hline 
\hline                
Parameter &  Best-fit value  \\  
\hline      
\hline
Half opening angle  of the precession cone & $\psi = 31 \pm 5\degr$    \\
&   \\
Inclination of the  jet precession axis  &  $i = 86 \pm 7\degr$      \\
with respect to the l.o.s.  &    \\
&   \\
Position angle of the jet precession axis      & $\chi = 134 \pm 6\degr$                  \\     
&   \\
Precession period                    &   $P = 1821 \pm 503$  d           \\
Precession phase\parnote{In fitting the model to the data, we implicitly assumed that the variation of the precession phase was negligible over the three days of observation, which is consistent with the 1821 days precession period obtained. Therefore, the precession phase provided here corresponds to an average over our observing days.} & $\phi_{\rm prec} = 1.1 \pi \pm 0.3 $     \\
Jet velocity                                                & $v_{\rm jet} = (0.5 \pm 0.2) c$            \\
Distance                                                    & $d = 9.4 $ kpc (fixed)         \\
 \hline
 Approaching jet (N)                                   & $s_{\rm jet} = +1 $   (fixed)    \\
Receding jet (S)                                        &  $s_{\rm jet} = -1 $    (fixed)    \\
Sense of rotation  (counterclockwise)    &   $s_{\rm rot} = +1$   \\
 \hline       
 \hline
\end{tabular}
\parnotes
\end{table}

\begin{figure}
\includegraphics[width=0.49\textwidth]{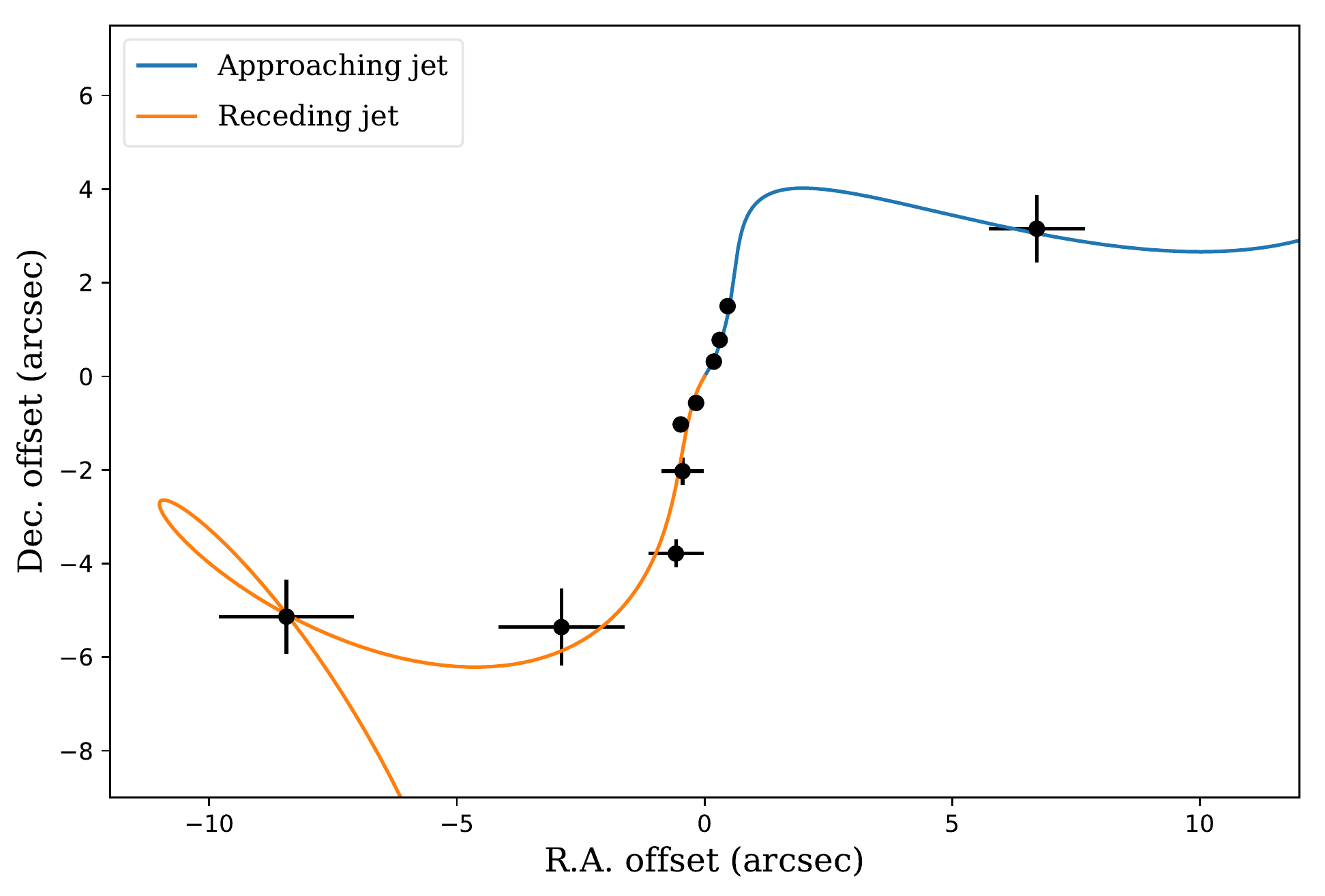} 
\caption{Positions of the jet components, plotted as R.A. and Dec. offsets from the core, together with the best fit precessing jet model whose parameters are given in \autoref{param}. The approaching jet (North-West) is represented in blue and the receding jet (South-East) in orange. Note that ``approaching'' and ``receding'' refers here to the orientation of the jet precession axis w.r.t. the line of sight. However, this axis is oriented at 86 degrees to the line of sight and given the half-opening angle of 31 degrees of the precession cone, each jet component will switch from approaching to receding, and vice versa, every precession cycle.}
\label{fit1}
\end{figure}

\begin{figure*}
\includegraphics[width=0.85\textwidth]{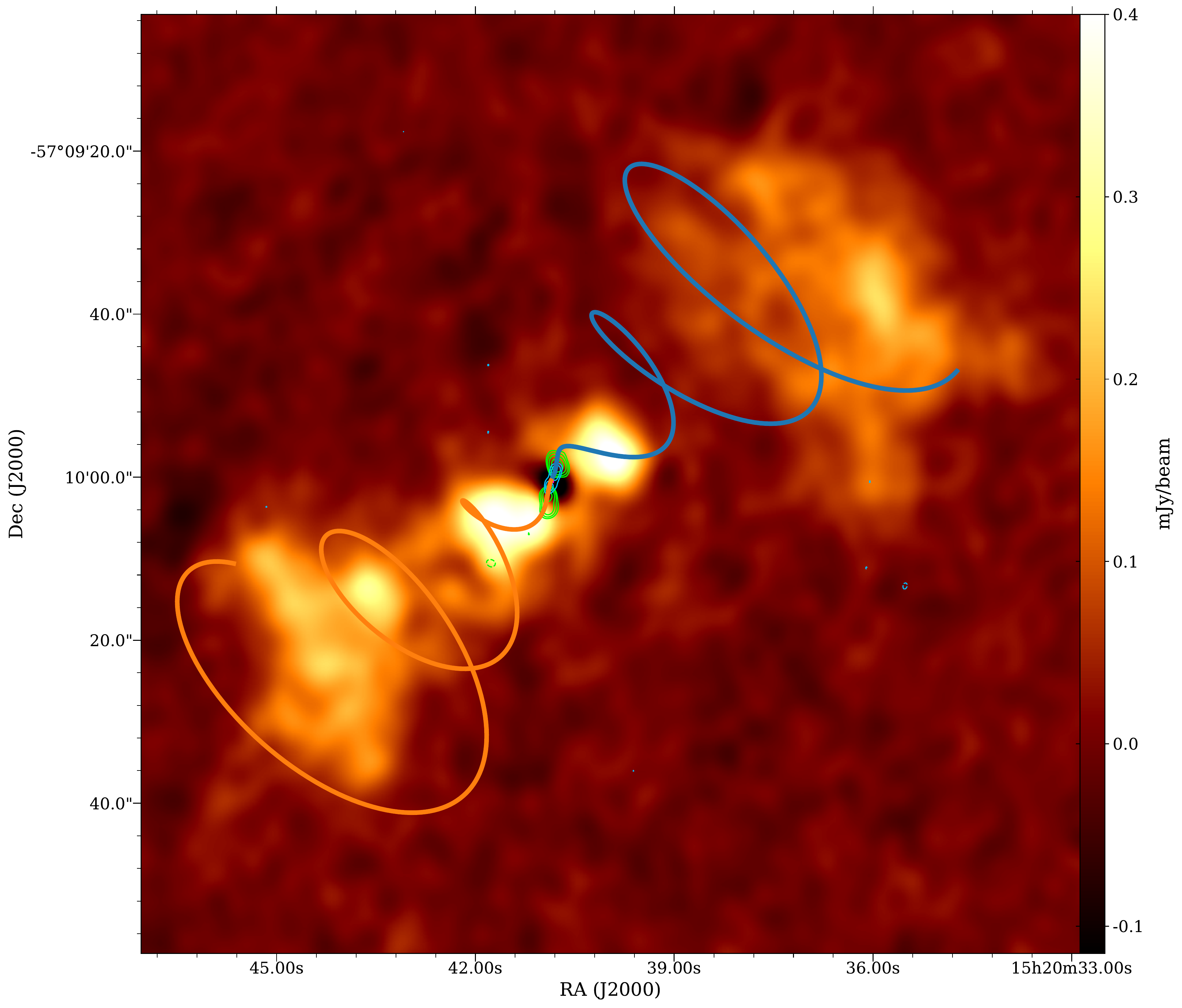} 
\caption{Precessing jet model overlaid on the 2.1 GHz image and the 5.5, 9.0 and 34 GHz contours. Only the data within the central 20 arcsec were used for fitting (see \autoref{fit1}). We here show the extrapolation of the model up to the distances of the lobes to verify consistency of the model with the orientations and angular sizes of the lobes. The approaching jet (North-West) is represented in blue and the receding jet (South-East) in orange. Note that ``approaching'' and ``receding'' refers here to the orientation of the jet precession axis w.r.t. the line of sight. However, this axis is oriented at 86 degrees to the line of sight and given the half-opening angle of 31 degrees of the precession cone, each jet component will switch from approaching to receding, and vice versa, every precession cycle.}
\label{fit2}
\end{figure*}

\vspace{0.5cm}

We obtain the best fit with the parameters listed in \autoref{param}. The associated uncertainties have been obtained using a Monte Carlo Markov Chain (MCMC) analysis and correspond to a $1\sigma$ quantile, estimated as half the difference between the 15.8 and 84.2 percentiles. Note that we fixed the distance to 9.4 kpc while \citet{heinz15} provide an error range of $9.4^{+0.8}_{-1.0}$ kpc. These uncertainties will primarily impact the estimated jet velocity. To measure this impact we ran an MCMC analysis fixing all the parameters to their best-fit value except velocity and distance, the latter being free to vary between 8.4 kpc and 10.2 kpc. We derive a $1\sigma$ error of $0.06 c$ on the velocity due to the uncertainty on the distance. The error bar on the velocity provided in \autoref{param} takes into account this additional source of uncertainty. 

Data points and jet model are plotted in \autoref{fit1} while \autoref{fit2} shows the model overlaid on the radio contours and image. The model depicts a source with mildly relativistic jets ($v\sim 0.5 c$), inclined close to the plane of the sky ($i\sim 86\degr$) and precessing over a period of approximately 5 years. Despite being simple, as it assumes continuous ejection at constant velocity (which is likely not the case), the model successfully reproduces the jet flow path. This is obviously not a proof that precession is the correct interpretation nor that the set of parameters we find is unique. This essentially demonstrates that precession is a valid interpretation. As a consistency check, using our best-fit parameters, we calculated the jet position angle and the north-to-south jet flux ratio predicted by the model if we were observing at milli-arcsecond (mas) scale on 2010 July 28 (507 days before our observations), the date of the LBA observations of \citet{miller-jones12}. At a distance of 10 mas from the core,  the model predicts a position angle of $105 \pm 8\degr$ and a jet flux ratio of $0.60 \pm 0.26$ which is in global agreement with the position angle of $112 \pm 1.5 \degr$ and the jet flux ratio of $0.71 \pm 0.11$ measured\footnote{There is no uncertainty reported on the jet position angle in \citet{miller-jones12}. We estimated an error of $1.5\degr$ on the jet angle based on the reported PSF size and  signal-to-noise ratio. The jet flux ratio and associated error has been estimated from the contours and the rms noise level of their Figure 2b.} by \citeauthor{miller-jones12}

Finally, it is interesting to note that Cir X-1 has been recently proven to be the youngest known X-ray binary by constraining the age of its surrounding supernova remnant \citep{heinz13}. The only other  established Galactic X-ray binary with a supernova remnant is the black hole candidate SS433 \citep{geldzahler80}. Indications that jets could also be precessing in Cir X-1 add further similarities between the two sources. It suggests that precession is related to the young age of these systems where alignment of the spin of the two components with the orbital axis has not been achieved yet. Furthermore, the mass transfer may not have started immediately after the supernova, further reducing the time available for spin-orbit alignment. (see e.g. \citealt*{brandt95,martin08} and discussion in \citealt{heinz13}).  

\subsection{Precession period}\label{sec:prec-period}

The precession period we obtain seems particularly long when compared with precession periods observed in black hole X-ray binaries which are typically of the order of tens to few hundred days (e.g. 162.5 days for SS433, \citealt{eikenberry01}, or 485 days for 1E 1740.7--2942, \citealt{luque-escamilla15}). To our knowledge, there is no reported jet precession period for a (confirmed) neutron star X-ray binary we could straightforwardly compare our results with. The nature of the compact object could be the origin of this notable difference in precession periods between \cirx\ and the BHXBs.
We note however that \citet*{rajoelimanana11} reported superorbital periodicities in Be X-ray binaries in the Small Magellanic Cloud, some of which were greater than 1000 days, for orbital periods of a few tens of days. The authors attribute these long-term variability to properties of the Be star disk, though. While the nature of the companion star is still unclear in \cirx, the possibility of a Be star is not favoured due to the lack of double-peaked lines in optical and infrared spectra and an orbital period shorter than any other Be/X-ray binaries \citep[see discussion in][]{johnston16}.

\begin{figure}
\includegraphics[width=0.47\textwidth]{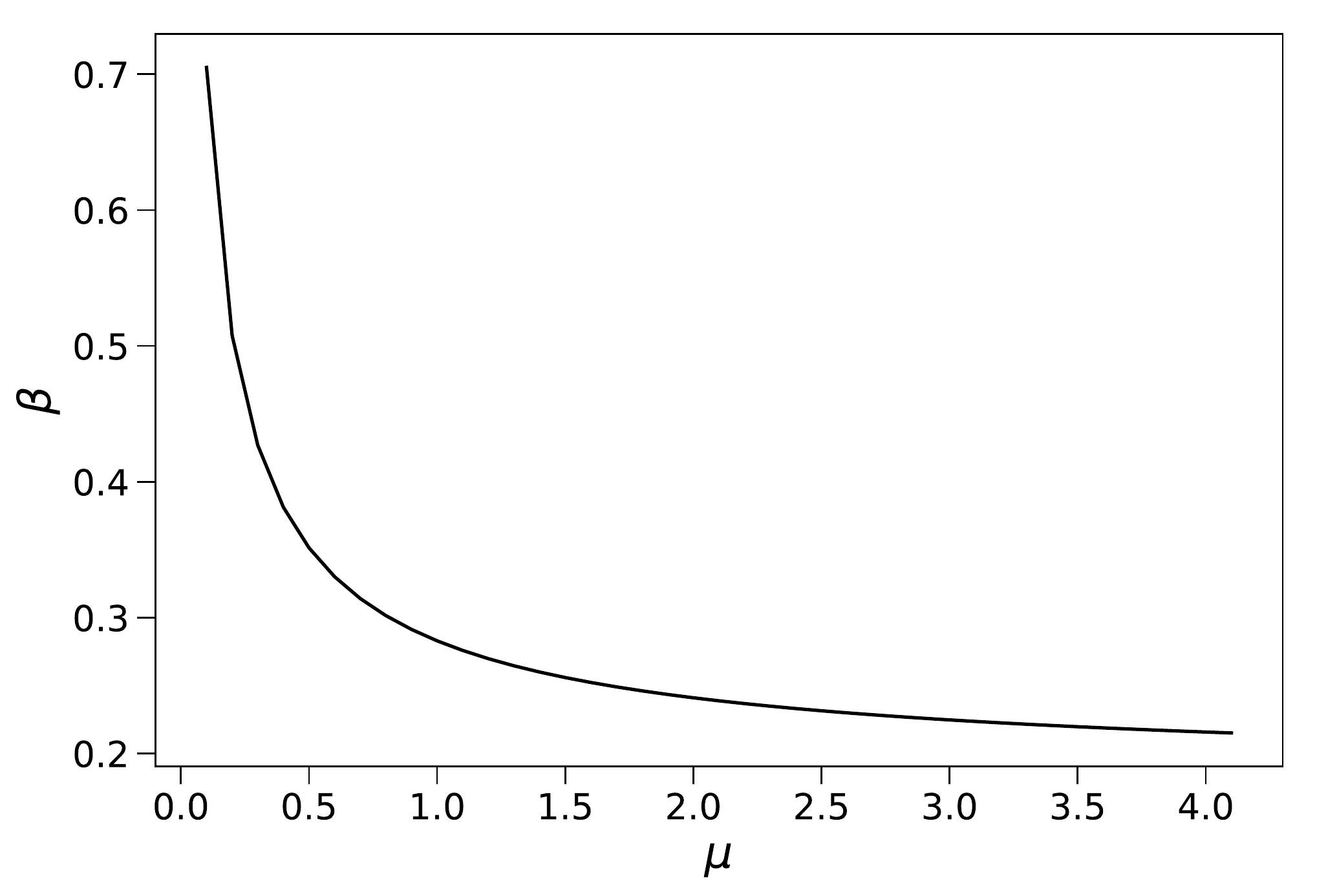} 
\caption{Expected accretion disc outer radius (given as a fraction $\beta$ of the Roche lobe radius) as a function of the mass ratio $\mu$ in the hypothesis that precession of the accretion disc is induced by tidal interactions with the companion star. See \autoref{eq-tidal} and text in Section~\ref{sec:prec-period} for details.}
\label{betamu}
\end{figure}

From the current literature, it is not obvious that a 5-year precession period could be explained by the theory of an underlying warped accretion disc driving the precession of the jets. Although it is beyond the scope of this paper to investigate this aspect in details, some simple estimates can be made. As an example, \citet{massi10} examine two mechanisms, tidal and Lense-Thirring precessions, to explain the precession of the jets in the gamma-ray binary LS I $+61\degr303$, a compact object on an eccentric orbit around a Be-type star (\citealt*{taylor82}; \citealt{casares05}). As stated in \citeauthor{massi10}, precession of an accretion disk could result from the misalignment of the compact object rotational axis with respect to the orbital plane; a misalignment produced by an asymmetric supernova explosion of the progenitor. Then two resulting configurations can be considered: either the disc is mostly coplanar with the rotational plane of the compact object and would therefore be subject to the gravitational torque of the companion star, or the disc is mostly coplanar with the orbital plane and tilted w.r.t the compact object, which would induce Lense-Thirring precession. The latter mechanism can be excluded in our case as it produces only  precession periods shorter than a few days for a large range of parameters \citep[see][]{massi10}. On the other hand, tidal precession leads to precession periods of a few hundred days or longer which is more consistent with our results. We can thus try and put constraints on the system configuration assuming tidal precession is the driving mechanism. Following \citet{larwood98}, we can write the accretion disc outer radius as a fraction $\beta$ of the Roche lobe radius. Then, combining Eq. 4  from \citet{larwood98} with the analytical expression of the Roche radius given by \citet{eggleton83}, we can write the expected precession period as follow: 
 
 \begin{equation}\label{eq-tidal}
 P_{\rm tidal} = \frac{7}{3} \frac{P_{\rm orb}}{\cos \delta} \frac{(1+\mu)^{1/2}}{\mu} \left( \frac{0.49 \beta}{0.6 + \mu^{2/3} \ln(1 + \mu^{-1/3})} \right)^{-3/2}, 
 \end{equation}
where $P_{\rm orb}$ is the orbital period, $\delta$ is the inclination angle of the accretion disc with respect to the orbit and $\mu = \frac{M_{\rm \star}}{M_{\rm neutron \, star}}$ is the mass ratio. As mentioned earlier, the nature of the companion star is uncertain. Possibilities range from a low-mass companion \citep{johnston16} to a B5 supergiant \citep{jonker07}. Taking $M_{\rm neutron \, star} = 1.4 M_{\sun}$, we must therefore consider the range $0.1-4.2$ for the mass ratio. In addition, assuming the jet precession axis is orthogonal to the accretion disc plane implies  $\delta = \psi = 31 \degr$, the half opening angle of the precession cone in our model. Using these parameters together with $P_{\rm orb} = 16.6$ days and $P_{\rm tidal} = 1821$ days, we plot in \autoref{betamu} $\beta$ as a function $\mu$. The figure shows that a 5 year precession period can be achieved for $\beta$ values between 0.2 and 0.7 depending on the mass ratio. These values are a bit smaller than the standard Paczy\'nski radius of an accretion disc for this mass ratio range \citep[i.e. $\beta = 0.7-0.86$, ][]{paczynski77}. This would indicate that only a portion of the disc is warped out, which could be expected since matter should initially flow from the companion with an angular momentum aligned with the binary orbit. We can conclude that tidal precession is a plausible mechanism to explain our long precession period although the above calculations assume a circular orbit. The significant eccentricity of \cirx\ orbit should be properly taken into account to derive meaningful values of the warped accretion disc size.

Two additional mechanisms are commonly invoked to explain warping and precession in accretion discs, namely, radiation-driven \citep*{pringle96,pringle97,maloney96,maloney98} and magnetically-driven precession \citep*{lipunov80,lai99,lai03,terquem00}. However, these two mechanisms rely on a significant number of unconstrained parameters related to the micro-physics of the accretion disc and/or the magnetic field topology and intensity. This lack of knowledge leads to estimated precession periods ranging from $10^{-4}$ to $10^{5}$ years in the case of neutron star X-ray binaries \citep[see][]{caproni06}. In this context, a 5-year precession period does not seem unreasonable.

Tidal, radiatively-driven and magnetically-driven precessions therefore appear as possible mechanisms to explain our results and deserve further investigations to conclude on their applicability to \cirx . 

\subsection{Orientation of the system}

Regarding the inclination of the jet precession axis w.r.t the line of sight, we note that the value of $86\degr$ that the model predicts is not consistent with the upper limit on the jet viewing angle of $\la 5\degr$ reported by \citet{fender04a} based on high Lorentz factors inferred from the time delay between radio core flaring and re-brightening of the downstream material in October 2000. Aside from this inconsistency with the fitted parameter (which might be due to wrong assumptions and/or wrong modelling on our side), for such a low inclination to be compatible with the jet flux ratio, the actual velocity of the jet components must be $\la 0.1 c$. Such a low velocity would be in turn inconsistent with the proper motions of 16 and 35 mas d$^{-1}$ observed by \citet{miller-jones12}. It is therefore difficult to reconcile the properties of the system inferred from observations in 2000 with the properties of the jets observed in 2010 and 2011. A possibility would be that the system underwent a significant shift in orientation. Modelling of W50's structure \citep{goodall11} has shown that it required at least 3 distinct epochs of outflows from SS433, each with different properties (such as precession angle) to produce the visible structure, rather than any one stable configuration. Thus, a significant shift in Cir X-1's outflow properties is not a radical concept.   

Another possibility would be that the downstream lobe flare observed in 2000 October 21 is in fact associated to a core flare from an earlier periastron passage than the one immediately before. We can check if our model parameters are consistent with this hypothesis. Let us first extend our precession model backward in time as we did to compare with \citet{miller-jones12} observations. On 2000 October 21 and at a distance of 2.3 arcsec from the core, the model predicts a position angle of $136 \pm 25\degr$ and a  North-to-South jet flux ratio of $0.18 \pm 0.12$. From coordinates and flux densities given in \citet{tudose08}, the observed jet position angle is $114 \pm 10\degr$ and the jet flux ratio $0.16 \pm 0.03$, therefore broadly consistent with the model predictions. Assuming that our modelling is correct, we derive an instantaneous inclination of the jets w.r.t the line of sight of $56 \degr$. For such an inclination, the time needed for a shell of matter or a shock to travel 2.3 arcsec at $0.5c$ at a distance of 9.4 kpc is 216 days which corresponds to 13 orbital periods. Taking now into account the uncertainty on the velocity and distance, we obtain a range of travel time from 8 to 24 orbital periods. Given this rather large range it is difficult to conclude on this possible interpretation. Again here further investigations and a more precise modelling would be needed.

Another orientation-related remark is that the combination of the inclination $i$ of the system w.r.t. the line of sight and the opening angle $\psi$ of the precession cone implies that at different times, one side of the jet will be approaching or receding, which may explain the shift in north:south jet brightness ratios reported by \citet{calvelo12}.

\subsection{Core variability}\label{sec-corevar}

\begin{figure}
\includegraphics[width=0.50\textwidth]{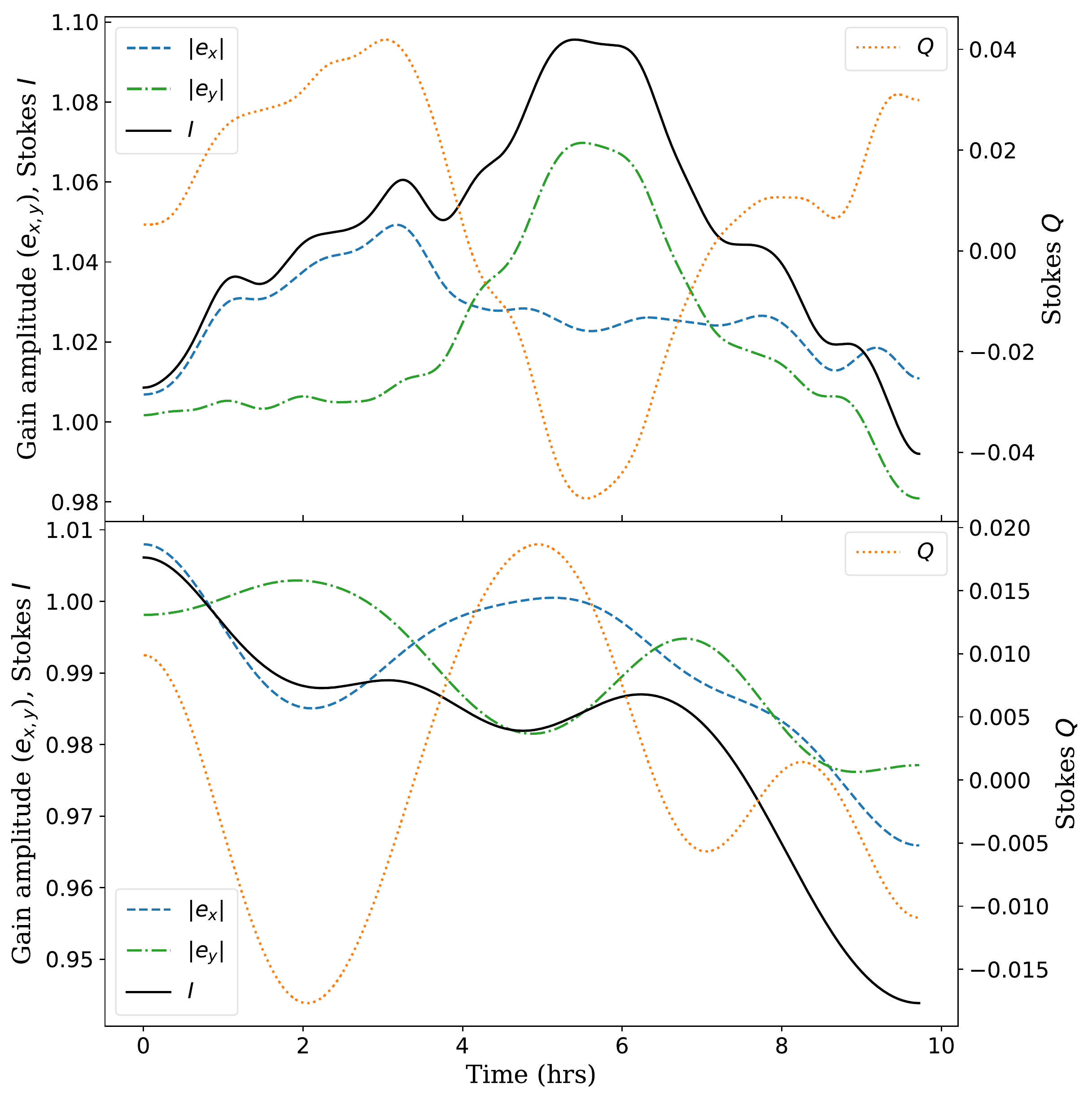} 
\caption{Differential gain-amplitudes as a function of time computed during the calibration of the 5.5 GHz (top panel) and 9.0 GHz (bottom panel) visibilities (see text in Section~\ref{sec-corevar}). The dashed blue and dash-dot green lines represent the gain-amplitudes for the X and Y correlations respectively. The solid black line corresponds to the apparent Stokes I flux relative to the model flux (i.e. $I_{\rm app}/I_{\rm mod}$, see \autoref{eq-Iapp}). The dotted orange line corresponds to the apparent Stokes Q flux relative to the model flux (i.e. $Q_{\rm app}/I_{\rm mod}$, see \autoref{eq-Qapp}). }
\label{dE}
\end{figure}

For the data reduction presented in Section~\ref{datared_5g} we introduced a differential gain term into the measurement equation to absorb the potential variability of the unresolved emission from the binary core during the course of the observation.

As such, the differential gain-amplitudes encode the light-curve of the source since other corrupting effects affecting the entire field should have been taken care of by the global gain solutions. If the target source model is unpolarised (which is our case), with constant flux $I_{\rm mod}$, and the differential gain solutions are given by a diagonal matrix with $e_x(t), e_y(t)$ on the diagonal (corresponding to the X and Y polarisations of the antennas' receivers), then the apparent Stokes I flux is given by: 

\begin{equation}\label{eq-Iapp}
I_{\rm app} = I_{\rm mod} \; \frac{\left(|e_x|^2 +|e_y|^2\right)}{2}
\end{equation}
and the apparent Stokes Q flux by:
\begin{equation}\label{eq-Qapp}
Q_{\rm app} = I_{\rm mod} \frac{\left(|e_x|^2 - |e_y|^2\right)}{2}
\end{equation}
(we only processed dual-polarised data for this study, so U and V are irrelevant).

 \autoref{dE} shows the differential gain-amplitudes as a function of time, as well as the apparent $I_{\rm app}$ and $Q_{\rm app}$ fluxes (right axis), relative to the model flux $I_{\rm mod}$. The solutions were obtained using a common $\Delta E$ term for all antennas, using the smooth solution mode, with a Gaussian weighting kernel of 15 min and 30 min for the 5.5 and 9.0 GHz data respectively (a larger kernel was chosen for the 9.0 GHz dataset due to lower signal-to-noise ratio).  We first note the difference between the $I_{\rm app}$ fluxes at 5.5 and 9.0 GHz. Although the data at these two frequency bands have been acquired simultaneously, the corresponding light-curves are not identical. While the 5.5 GHz flux rises and fades during the observation, the 9.0 GHz one shows a general decrease. This could be related to the fact that we are probing different structure sizes at 5.5 and 9.0 GHz and/or by spectral variations of the jets.  A second notable feature appears from these figures. For both frequency bands, we note significant variations of the $Q_{\rm app}$ fluxes which should not be present if the source was unpolarised. Indeed, since the ATCA is an alt-azimuth mount telescope, the X and Y orthogonal feeds of each antennas rotate with respect to the equatorial frame. This causes the actual response of linearly polarised feeds to vary with the angle $\chi$ between the ``sky'' and the feed  (which varies with time), according to:
 \begin{eqnarray*}
XX = I + Q \cos(2\chi) + U \sin(2\chi) \\
YY = I - Q \cos(2\chi) - U \sin(2\chi)
\end{eqnarray*}
Consequently, given that our source model is unpolarised, if the true source is actually linearly polarised ($Q \neq 0$ and/or $U\neq 0$), the differential gain solutions will correct for these unmodeled variations. Therefore $Q_{\rm app}$, as we defined it, should be proportional to  $Q \cos(2\chi) + U \sin(2\chi)$ which explains the oscillations we see  on \autoref{dE}. The fact that \cirx\ emission is polarised is not surprising and consistent with the synchrotron optically thin spectrum we obtain based on our core flux densities measurements (see \autoref{spec} and section \ref{sec-spec}). Finally, we note that the $Q_{\rm app}$ curves at 5.5 and 9.0 GHz are not similar and even seem anti-correlated. The only reasonable explanation that we see is a $\sim 90\degr (\pm n\pi) $ Faraday rotation induced by the local and/or interstellar medium. Such a rotation angle implies a lower limit on the rotation measure of $\gtrsim 8 \times 10^{-2} \text{cm}^{-2}$ which in turn sets a lower limit on the average of the magnetic field parallel to the line of sight, weighted by the electron density: $n_e B_{\|} \gtrsim 10^{-7} \text{Gauss cm}^{-3}$ for a distance of 9.4 kpc.

\subsection{Spectrum and nature of the unresolved central outflow}\label{sec-spec}

\begin{figure}
\centerline{\includegraphics[width=0.50\textwidth]{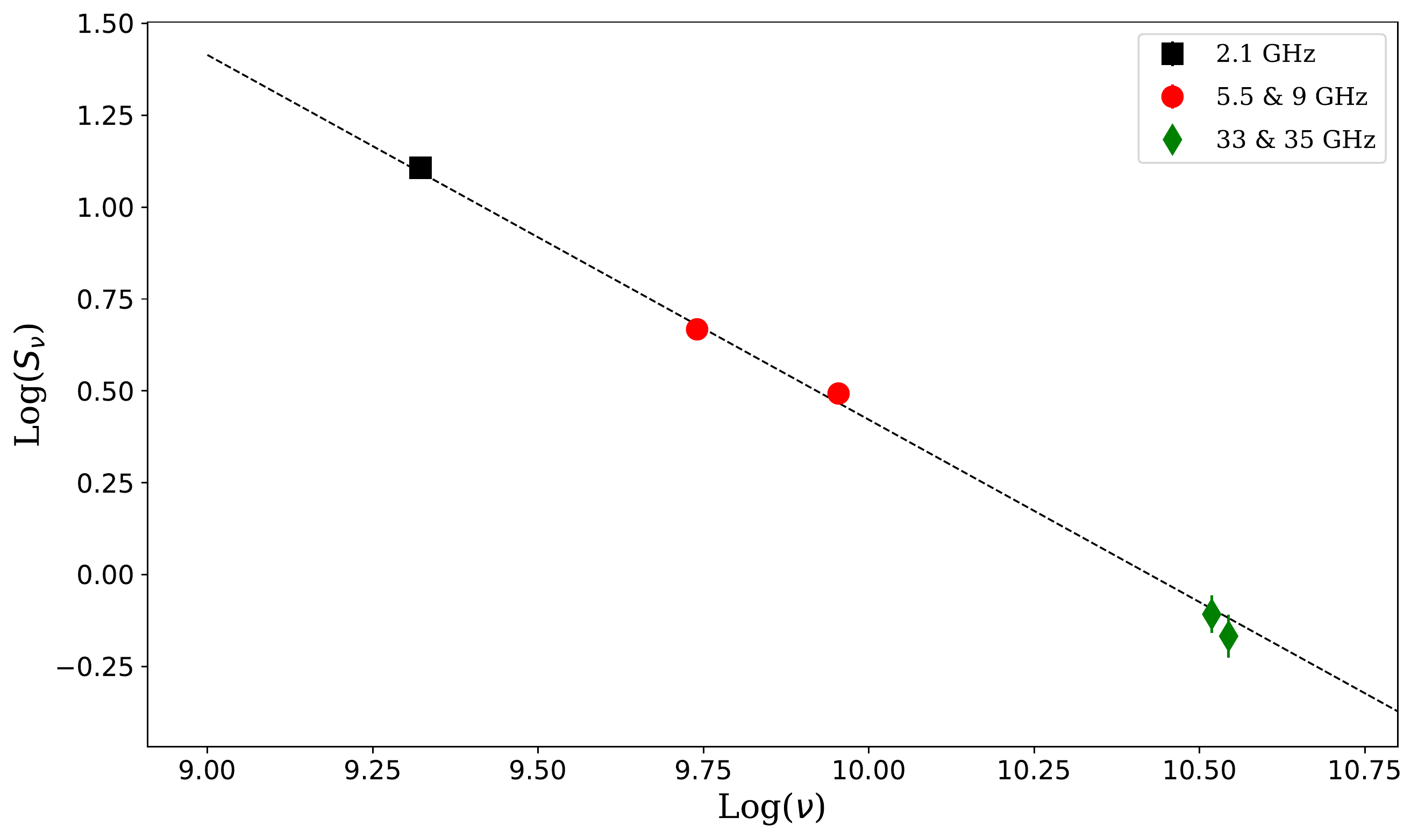}}
\caption{Cir X-1 spectrum of the unresolved core. Flux densities and uncertainties are plotted and obtained from point source fitting to the image plane (see \autoref{table}). Fitting a power-law to the data we derive a spectral index of $- 0.99 \pm 0.02$ .}
\label{spec}
\end{figure}

From the point source fitting to the unresolved core we performed on each dataset, we obtain the flux density measurements summarised in \autoref{table} with the corresponding spectrum shown in \autoref{spec}. A power-law fit to the data gives a spectral index of $- 0.99 \pm 0.02$, consistent with optically thin synchrotron emission. Note that the steepness of the spectrum is likely accentuated by the presence of jet structures at all scales that are unresolved to the lower frequency beams.

\citet{migliari10} have shown that the available spectrum of the atoll-type NSXB 4U 0614+091 can be fitted by a flat power-law from the radio to the mid IR implying emission originates in a compact optically thick jet as observed in black hole XRBs in the hard state. However, the jet break frequency, i.e. the frequency at which the jet becomes transparent to its own synchrotron emission, was lower by a factor of $\sim10$ than the one measured in the black hole GX 339--4 during a fully established hard state \citep*{corbel02}. Other atoll sources detected in radio have only shown optically thick synchrotron emission consistent with the presence of compact jets (\citealt{rutledge98, moore00, migliari04}; \citealt*{migliari06}; \citealt{miller-jones10}).
Our results indicate that the radio to mm spectrum of  Cir X-1 is optically thin during non-flaring periods when, presumably, compact jets could form. This highlights the existing behavioural differences between atoll sources and this NS system. Cir X-1 has shown evidence for discrete jet ejecta moving away from the core following flare events: behaviour which atoll sources lack. If parallels can be made between NS and BH systems, then these flares and subsequent ejections should coincide with periods of high accretion onto the neutron stars. Although Cir X-1 is a special case within the standard classification framework, it is believed to have one of the highest accretion rates among the NS systems being regularly above the Eddington limit \citep{heinz15}. A characteristic shared with Z-sources for which optically thin radio cores are also observed \citep{migliari11}. More generally, our results are consistent with the unified picture proposed to explain the different sub-classes of NSXBs as being mainly driven by mass accretion rate (see e.g. \citealt{homan10}, \citealt*{fridriksson15}). The weakly accreting atoll sources, producing only compact jets, would be the equivalent of BH in the hard state. When producing jets, the highly accreting NS systems  would correspond to the intermediate state (hard-to-soft) BHs displaying mainly transient ejections (see \citealt{migliari06, migliari11} and Motta \& Fender, submitted,  for a detailed discussion on jets in NS systems and comparison with BHs).

\section{Summary}\label{sum}

We have described results of ATCA multi-frequency radio observations of the accreting neutron star Circinus X-1. We presented advanced data processing methods that we used to overcome multiple issues due to intrinsic variability of the source and direction dependent effects. We produced what we believe are the most detailed images so far of the jet-SNR interaction in this source and we think these images could be exploited more, in the future, by groups modelling jet-media interaction.

The centimetre and millimetre radio maps revealed symmetrical jet structures with significant variation of the jet axis angle. We explored the possibility that such a morphology would result from precession by modelling the jet flow path with a kinematic jet model. The result of the modelling suggests that precession is a plausible interpretation and implies that the source displays mildly relativistic jets seen at high inclination with respect to the line of sight. We discussed the incompatibility of our results with previous findings of highly relativistic jets seen at low inclination and concluded that a significant shift of the global orientation of the system might have happened over the last decades. 

We then analysed the differential gain solutions obtained during calibration of the centimetre datasets and recovered the intrinsic variability of the source during our observation. Power-law fits to the centimetre and millimetre flux densities of the source then indicated that optically thin synchrotron emission dominates within the extent of the point source regions. The outflowing modes of \cirx\ thus appear to be dominated by recurrent discrete ejection events in contrast with the compact jet state observed by e.g. \citet{migliari10} in 4U 0614+091.  This highlights the main differences between Atoll and Z-type NSXB and is consistent with the unified picture where mass accretion rate is the leading parameter behind the NSXB classification framework.

\section{Acknowledgements}
We would like to thank the referee for his/her constructive comments that helped improving the quality of the paper.
MC acknowledges the financial assistance of the National Research Foundation (NRF) of South Africa through an SKA SA Fellowship.
The Australia Telescope Compact Array is part of the Australia Telescope funded by the Commonwealth of Australia for operation as a National Facility managed by CSIRO. 
This research made use of APLpy, an open-source plotting package for Python \citep{robitaille12} and LMFIT, a non-linear optimisation package for Python (\url{https://lmfit.github.io/lmfit-py/}).




\bibliographystyle{mnras}
\bibliography{/Users/mickael/TAFF/Biblio/biblio}




%
%


\bsp	
\label{lastpage}
\end{document}